\documentstyle[aps,multicol,prb,epsf]{revtex}

\newcommand{\nil}{\hspace*{0em}}
\begin{document}
\draft
\title{The magnetic structure of the Jahn-Teller system LaTiO$_{\mathbf{3}}$}
\author{Robert Schmitz,$^1$ Ora Entin-Wohlman,$^2$ Amnon Aharony,$^2$ A.\,Brooks Harris,$^3$
and Erwin M\"uller-Hartmann$^1$}
\date{\today}
\address{$^1$Institut f\"ur Theoretische Physik,
Universit\"at zu K\"oln, Z\"ulpicher Stra{\ss}e 77, 50937 K\"oln, Germany\\$^2$School of
Physics and Astronomy, Raymond and Beverly Sackler Faculty of Exact Sciences,
\\Tel Aviv University,
Tel Aviv 69978, Israel\\$^3$Department of Physics and Astronomy, University of Pennsylvania,
Philadelphia, Pennsylvania 19104, USA}
\maketitle

\begin{abstract}
We investigate the effect of the experimentally observed
Jahn-Teller distortion of the oxygen octahedra in LaTiO$_3$ on the
magnetic exchange. We present a localized model for the effective
hopping between nearest-neighbor Ti ions and the intra-site
Coulomb interactions, based on a non-degenerate orbital ground
state due to the static crystal field. The latter corresponds to
an orbital order which recently has been confirmed experimentally.
Using perturbation theory we calculate, in addition to the
Heisenberg coupling, antisymmetric and symmetric anisotropy terms
of the superexchange spin Hamiltonian which are caused by the
spin-orbit interaction. Employing our superexchange Hamiltonian,
we deduce the magnetic order at low temperatures which is found to
be in good agreement with experiment.\\[3ex]
PACS numbers: 71.10.--w, 71.27.+a, 71.70.Ch, 75.10.Dg, 75.25.+z, 75.30.Et
\end{abstract}

\begin{multicols}{2}

\section{Introduction}

In the seventies, the orthorhombic perovskite LaTiO$_3$ was
considered as a typical example of an antiferromagnetic Mott
insulator ($T_{\text{N}}=146\,\text{K}$). \cite{imada1} The ground
state of the Ti ion is trivalent with a single electron in the $d$
shell. Recently, LaTiO$_3$ has attracted attention when an ordered
magnetic moment of 0.46\,$\mu_{\text{B}}$ was reported.
\cite{meijer} This value is surprisingly small for a single
electron with quenched orbital moment, for which one would have
expected   1\,$\mu_{\text{B}}$. In LaTiO$_3$, this value may be
reduced  by about 17\,\% due to quantum fluctuations of the 3D
Heisenberg model, and by about further 14\,\% due to the on-site
spin-orbit coupling, (in conjunction with the crystal field, see
below), leading to an overall estimate of 0.72\,$\mu_{\text{B}}$.

A previous attempt to explain this unexpected finding has
neglected  the Jahn-Teller distortion of the oxygen octahedra in
LaTiO$_3$  and assumed that the symmetry of the unit cell is
strictly cubic. In such a situation the $t_{2g}$ ground state of
the Ti ion is three-fold degenerate and the orbital moment is
unquenched. Hence, it was proposed to consider LaTiO$_3$ as an
orbital liquid in order to explain the reduction of the ordered
moment by orbital fluctuations. \cite{khaliullin}

However, recent experiments give a strong indication of the
importance of the Jahn-Teller distortion in LaTiO$_3$, and in
particular enable, using recent structural data, to estimate  the
splitting it induces in the $t_{2g}$ levels: There is a
crystal-field gap of about 0.2\,eV between the non-degenerate
ground state and the next excited level. \cite{cwik}  This value
 has been confirmed by a study of photoelectron spectroscopy
\cite{haverkort} and is at least one order of magnitude higher
than any superexchange energy in LaTiO$_3$. \cite{keimer}
Consequently, the orbital order at low temperatures  is not
induced by the superexchange. Rather, the orbital degree of
freedom is frozen by the crystal field. The scenario of suppressed
orbital fluctuations has also been confirmed by a recent LDA+DMFT
study. \cite{pavarini} The assumption of Ref.
\onlinecite{solovyev} that the crystal-field splitting, the
superexchange, and the spin-orbit coupling are all of the same
order is inconsistent with the photoelectron spectroscopy of Ref.
\onlinecite{haverkort}. In addition, a higher value,
0.57\,$\mu_{\text{B}}$, of the ordered moment has been recently
reported, \cite{cwik} making the discrepancy between experiment
and theory (which gives 0.72\,$\mu_{\text{B}}$) even smaller.

The Jahn-Teller effect in LaTiO$_3$ is caused  by the  twisting of
the Ti--O bonds with respect to each other (i.\,e. by differences
between the O--O bond lengths), rather than by differences between
the Ti--O bond lengths.  The non-degenerate ground-state orbital
due to crystal-field calculations given in Ref. \onlinecite{cwik}
is consistent with the orbital order found in NMR measurements of
the Ti--3$d$ quadrupole moment. \cite{kiyama} The presence of
orbital order at low temperatures has been also concluded from
measurements of the dielectric properties and the dynamical
conductivity. \cite{lunkenheimer} As opposed to these findings, an
orbital contribution to the specific heat, which is predicted by
the orbital-liquid model, has not been found in experiment.
\cite{fritsch}

Hence, from the recent experiments it must be concluded that the
orbital-liquid  model is inappropriate for LaTiO$_3$. Moreover, it
has been proven, by exact symmetry arguments, that due to a
hidden symmetry the superexchange Hamiltonian used in Ref.
\onlinecite{khaliullin} cannot reproduce the observed magnetic
order of LaTiO$_3$. \cite{harris03,harris04}

Another model to explain the magnetic properties of LaTiO$_3$
proposed the lifting of the $t_{2g}$ degeneracy by the crystal
field resulting from the eight La ions which surround each TiO$_6$
octahedron---assuming undistorted octahedra, i.\,e. neglecting the
Jahn-Teller effect. \cite{imada3} This model predicts  a realistic
non-degenerate ground-state orbital for each Ti ion,
\cite{keimer,kiyama} and yields  plausible values for the
Heisenberg couplings between nearest-neighbor Ti ions.

However, there are two points missing in the calculations of Ref.
\onlinecite{imada3}. (i) The
crystal field due to the eight nearest La ions, which gives a
nearly equidistant splitting scheme between the three $t_{2g}$
orbitals, is only a first approximation of the full crystal field
due to all ions of the solid. It is preferable to
treat the electrostatic crystal field more accurately, employing
the Madelung sum. Such a treatment shows that the Jahn-Teller
effect leads to a non-degenerate $t_{2g}$ ground-state orbital and
two quasi-degenerate excited orbitals. \cite{cwik} (ii) Terms of
the exchange Hamiltonian which break the spin-rotational
invariance and cause the magnetic order have not been
considered.  Hence, the origin of the observed magnetic order is
not fully understood so far.

In the present paper, we investigate a model for the magnetism of
LaTiO$_3$, which starts from a point-charge calculation of the
static crystal field for the Ti ions via a full Madelung sum over
the crystal (as was already discussed in Ref. \onlinecite{cwik}).
Taking into account the recent structural low-temperature data and
using a Slater-Koster parametrization of the Ti--O hopping, we
calculate an effective hopping matrix between the $d$ orbitals of
nearest-neighbor Ti ions. Treating this Ti--Ti hopping and the
on-site spin-orbit coupling as perturbations, we calculate the
superexchange coupling between the non-degenerate crystal-field
ground states of the Ti$^{3+}$ ions. In treating the Ti$^{2+}$
ions, which appear as intermediate states of the exchange
processes, we take into account the full on-site Coulomb
correlations in terms of Slater integrals and diagonalize the
Coulomb Hamiltonian together with the crystal-field one.  The
spin-orbit coupling gives rise to antisymmetric and symmetric
anisotropies of the spin Hamiltonian. We calculate the isotropic
part of the exchange coupling and both kinds of the anisotropies
to leading orders. Using our exchange Hamiltonian, we determine
the classical ground state which gives the directions of the spins
in the ordered phase. The experimental data reveals  a G-type
antiferromagnetic order along the crystallographic $a$ axis, which
is accompanied by a small ferromagnetic moment along the $c$ axis.
\cite{cwik} Our calculation reproduces this  order. In addition,
we find a small A-type moment along the $b$ axis,  which  has not
yet been detected experimentally.

In the next section we present the details of our model. Section
III is devoted to the perturbation expansion yielding the
microscopic spin Hamiltonian. Section IV discusses the macroscopic
magnetic Hamiltonian and the resulting magnetic order of the
classical ground state. It includes as well a detailed comparison
with existing experimental data. Finally, we summarize our results
in Sec. V.

\section{The model}

\subsection{The crystal field}

\begin{figure}[htb]
\leavevmode \epsfclipon \epsfxsize=7.truecm
\vbox{\epsfbox{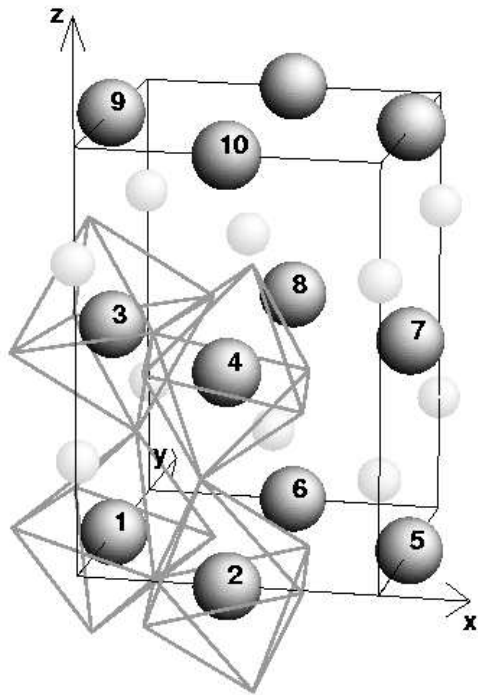}} \vspace{0.5cm}
 \caption{The crystallographic structure of LaTiO$_3$. The ten Ti ions, which
constitute the twelve inequivalent nearest-neighbor Ti--Ti bonds
are enumerated. For simplicity, oxygen octahedra are only shown
around the four crystallographically inequivalent Ti sites. La
ions from two layers are shown as small spheres. For example, the
sites 2 and 6 are crystallographically equivalent but the bond 61
emerges from the 12 bond by a glide reflection, so that the
effective hopping matrix for the bond 16 is different than that of
the 12-bond: It is the transposed one, see Table
\ref{effhop}.}\label{bonds}
\end{figure}

The unit cell of LaTiO$_3$ contains four Ti ions, see  Fig.
\ref{bonds} and Table \ref{par} (the Tables are given on pages \pageref{par}ff.),
and has the symmetry of the space
group $Pbnm$ (No. 62 in Ref. \onlinecite{hahn}). The symmetries of
this space group are listed in Table \ref{sym}. Given the position
of one La, Ti, O1, and O2 ion each (see Table \ref{par}), the
positions of all other ions in the unit cell follow from the
space-group symmetries. In order to use these symmetries
conveniently, we employ in our calculation the orthorhombic
orthonormal \cite{norm} basis for the Ti-$d$ orbitals
\begin{equation}
\big|xy\big>,\big|2z^2\big>,\big|yz\big>,\big|xz\big>,\big|x^2-y^2\big>,
\label{dbasis}
\end{equation}

\noindent
where the $x,y$ and $z$ axes correspond to the crystallographic
$a,b$ and $c$ axes.  In a cubic perovskite, the first two orbitals
would correspond to the $e_g$ orbitals and the three others to the
$t_{2g}$ orbitals. [Note that the pseudo-cubic basis for the $d$
orbitals, which is frequently used, is obtained from Eq.
(\ref{dbasis}) upon rotating the $x$ and $y$ axes by 45$^\circ$
around the $z$ direction.]

Using the structural data of Ref. \onlinecite{cwik} (taken at
$T=8\,\text{K}$), we have calculated the spectrum and the
eigenstates for the Ti ion located at  $(0,1/2,0)$, employing a
point-charge calculation of the static crystal-field Hamiltonian.
This calculation  uses the full Madelung sum over the crystal
(which is evaluated as an Ewald sum, see Appendix \ref{ewald}). It
requires the second moment, $\big<r^2\big>$, and the fourth
moment, $\big<r^4\big>$, of the effective ionic radius of the
Ti$^{3+}$-ion, see Appendix \ref{ewald}. We have used the values
$\big<r^2\big>=0.530 \,\stackrel{\mbox{\tiny$\circ$}}
{\mbox{\small{A}}}\!\!\nil^2$ and $\big<r^4\big>=0.554 \,
\stackrel{\mbox{\tiny$\circ$}}{\mbox{\small{A}}}\!\!\nil^4$.
\cite{altsh} The results  of the crystal-field calculation, which
are listed in Table \ref{cf},  exhibit a typical Jahn-Teller
$t_{2g}$ splitting scheme, where a non-degenerate ground state is
clearly separated from the two quasi-degenerate excited states.

\begin{figure}[htb]
\leavevmode \epsfclipon \epsfxsize=7.truecm
\vbox{\epsfbox{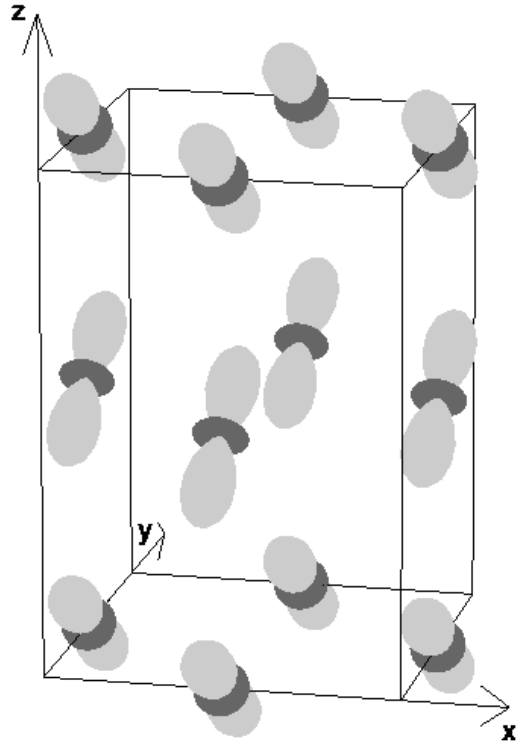}} \vspace{0.5cm}
 \caption{The orbital order of the Ti ions resulting from the
calculated crystal field (the energy scale of the superexchange
which could in principle also affect the orbital degree of freedom
is about one order of magnitude smaller than the crystal-field
gap). The order is ferro-orbital in the $ab$ planes and canted
antiferro-orbital between the planes.}\label{oo}
\end{figure}

The orbital order due to the static crystal field is shown in Fig.
\ref{oo}.  The ground-state orbital is given by
\begin{equation}
\big|0\big>=0.770\big|yz\big> \pm
0.636\big|x^2-y^2\big>.\label{ground}
\end{equation}

This state has approximately the  2$z^2$ structure in the
coordinate system in which the $y$ and $z$ axes are rotated by
$\pm 56^{\circ}$ around the $x$ axis.  \cite{cwik} The relative
sign of the linear combination alternates between neighboring $ab$
planes according to the mirror  planes at $z=1/4$, etc. In
the pseudo-spin language, \cite{kukho} we have ferro-orbital order
in the $ab$ planes and canted antiferro-orbital order between the
planes.  We note that this ground state is in perfect agreement
with experiment. \cite{kiyama} The ground state cited in Refs.
\onlinecite{kiyama} and \onlinecite{imada3} is given,  to a good
approximation, by
\begin{eqnarray}
\big|0'\big>&=& \underbrace{ \sqrt{\frac{2}{3_{\nil}}} }_{0.816}
\big|yz\big> \pm
\underbrace{\frac{1}{\sqrt{3}}}_{0.577}\big|x^2-y^2\big>. \label{ground2}
\end{eqnarray}
It practically coincides with our ground state,
\begin{equation}
\big|\big<0\big|0'\big>\big|^2=99.06\,\%. \label{coincidence}
\end{equation}

The ground state $\big|0\big>$ [see Eq. (\ref{ground})] of the
crystal field, which is occupied at each Ti site by a single
electron, is the starting point of our model. The perturbative calculation outlined
below is employed in order to evaluate
the magnetic superexchange coupling between Ti ions in this state.

\subsection{The  Hamiltonian} \label{modham}

The perturbation scheme is applied to a two-site cluster
consisting of two nearest-neighbor Ti ions, denoted by $m$ and $n$.
The unperturbed Hamiltonian acting on such a cluster is given by
\begin{equation}
H_{mn}^0=H_{mn}^{\text{cf}}+H_{mn}^{\text{c}}, \label{unperturbed}
\end{equation}
where $H_{mn}^{\text{cf}}$ is the static crystal field and
$H_{mn}^{\text{c}}$ describes the intra-ionic Coulomb correlations
of a doubly occupied $d$ shell. The perturbation calculation
requires a selected set of eigenstates and the corresponding
eigenenergies of $H_{mn}^0$. In our case, the eigenstates span a
Hilbert space which consists of two sectors. In the first, termed
the Ti$^{3+}$ sector, both Ti ions are trivalent. In the second,
called the Ti$^{2+}$ sector, one of the Ti ions is divalent (two
$d$ electrons on the same site) and the other is four-valent (an
empty $d$ shell). In the ground state of $H_{mn}^0$, which belongs
to the Ti$^{3+}$ sector,  both Ti ions are in the one-particle
ground state of $H_{mn}^{\text{cf}}$, modulo spin up or down on
each site. This leads to a four-fold degeneracy of the ground
state of the cluster. The complete Ti$^{3+}$ sector has 100 basis
states (made up of five orbital and two spin states on each of the two sites).
The Ti$^{2+}$ sector has 90 basis states (30 spin-triplet
states and 15 spin-singlet states of the doubly occupied $d$ shell
and a factor of 2 because each of the Ti ions  can be doubly
occupied while the other has an empty $d$ shell).

In order to calculate the spectrum of $H_{mn}^0$, we apply
$H_{mn}^{\text{cf}}$ on the Ti$^{3+}$ sector, and both
$H_{mn}^{\text{cf}}$ and $H_{mn}^{\text{c}}$ on the Ti$^{2+}$
sector. $H_{mn}^{\text{c}}$ is parametrized in terms of the Slater
integrals $F_2$ and $F_4$, \cite{slater} and the effective Ti--Ti
charge-transfer energy $U_{\text{eff}}$. This energy is the
difference between the four-fold degenerate ground state of the
cluster (which is the lowest level of the Ti$^{3+}$ sector) and
the lowest level of the Ti$^{2+}$ sector (where
$H_{mn}^{\text{cf}}$ and $H_{mn}^{\text{c}}$ are diagonalized
simultaneously). We use $F_2=8 F_4/5=8.3\,\text{eV}$ from an
atomic Hartree-Fock calculation, \cite{watson} and
$U_{\text{eff}}=3.5\,\text{eV}$ from the analysis of the
photoemission spectra and first-principles calculations.
\cite{saitoh}

The perturbation part of the Hamiltonian, $V_{mn}^{\nil}$,
consists of an effective Ti--Ti tunnelling term,
$H_{mn}^{\text{tun}}$, and the on-site spin-orbit interaction,
$H_{mn}^{\text{so}}$,
\begin{equation}
V_{mn}^{\nil}=H_{mn}^{\text{tun}}+H_{mn}^{\text{so}}.\label{perturbed}
\end{equation}
The tunnelling Hamiltonian is given in
terms of an effective hopping matrix, $t_{mn}$, between the $m$
and the $n$ Ti ions,
\begin{equation}
H_{mn}^{\text{tun}}=\sum_{ij} t_{mn}^{ij} d_{mi}^{\dag}d_{nj}^{\hspace*{0em}}
+\text{h.\,c.}, \label{Htun}
\end{equation}
where the indices $i$ and $j$ enumerate the eigen-orbitals of
$H_{mn}^{\text{cf}}$ (see Table \ref{cf}). The spin-orbit coupling
is given by
\begin{equation}
H_{mn}^{\text{so}}=\lambda \hspace*{-0.5em} \sum_{k=m,n} \hspace*{-0.5em}
{\mathbf{l}}_k{\mathbf{s}}_k,
\end{equation}
where ${\mathbf{l}}_k$ denotes the
angular momentum operator of the Ti ion at the $k$ site,
${\mathbf{s}}_k$ its spin operator, and $\lambda $ the
spin-orbit coupling strength. We use $\lambda=18\,\text{meV}$. \cite{imada3}

The 14\,\% reduction of the magnetic moment, alluded to in Sec. I
above, is obtained upon diagonalizing together
$H_{mn}^{\text{cf}}$ and $H_{mn}^{\text{so}}$ for a single
Ti$^{3+}$ ion. The reduction in the magnetic moment due to the not
fully quenched orbital moment is estimated
as follows. The spin along a certain selected direction does not
commute with $H_{mn}^{\text{so}}$ and therefore it is not a good
quantum number for the combined Hamiltonian
$H_{mn}^{\text{cf}}+H_{mn}^{\text{so}}$. However, the eigenstates
of this combined Hamiltonian are symmetric or antisymmetric with
respect to time-reversal. This leads to five Kramers doublets for
the single Ti$^{3+}$ ion. Since the ordered magnetic moment is
mainly of the G-type, and is oriented along the $x$ axis,
\cite{cwik} we use those doublets to find the expectation values
of the angular momentum. By choosing the largest possible
polarization of the magnetic moment along the $x$ axis out of all
the linear combinations of the ground-state doublet, we find the
expectation value
$\big<l^x_k+2s^x_k\big>\,\mu_{\text{B}}^{\nil}=0.86\,\mu_{\text{B}}^{\nil}$.
This effect is not included in our perturbation calculation of the
magnetic exchange. However, it does explain partially  why the
observed ordered moment along the $x$ axis is reduced with respect
to $1\,\mu_{\text{B}}$.

The dominant hopping process between two nearest-neighbor Ti ions
is mediated via the oxygen ion which is nearest to both of
them. Let $t^{i\alpha}_m $ be the hopping matrix element of an
electron in the $p$ orbital $\alpha$ on the oxygen ion  into
the $i$ state of the  Ti ion located at $m$. The
effective hopping between the Ti ions is then given by
\begin{equation}
 t_{mn}^{ij}=-\frac{1}{\Delta_{\text{eff}}}
 \sum_{\alpha} t^{i\alpha}_m
 t^{j\alpha}_n=t_{nm}^{ji}.\label{EFFTUN}
\end{equation}
Here,
$\Delta_{\text{eff}}$ is the charge-transfer energy, which is
required to put an electron from an O ion on a Ti ion, and
$\alpha$ denotes one of the three $p$ orbitals on the oxygen (in
orthorhombic coordinates),
\begin{equation}
\big|x\big>,\big|y\big>,\big|z\big>. \label{pstates}
\end{equation}

Using the structural data taken at $T=8\,\text{K}$, \cite{cwik}
together with elementary geometric considerations, the Ti--O
hopping matrix elements can be expressed in terms of the
Slater-Koster parameters $V_{pd\sigma}$ and $V_{pd\pi}$.
\cite{harrison} We use the values $V_{pd\sigma}=-2.4$\,eV,
$V_{pd\pi}=1.3$\,eV, and $\Delta_{\text{eff}}=5.5$\,eV,
\cite{imada3,saitoh} in conjunction with Eq. (\ref{EFFTUN}) to
compute the effective hopping matrices pertaining to the unit
cell. The results are listed in Table \ref{effhop}, which also
gives the symmetry properties of the hopping matrices between
different Ti--Ti bonds. The four inequivalent Ti sites of the unit
cell form twelve nearest-neighbor Ti--Ti bonds which are
crystallographically inequivalent, i.\,e. they do not evolve from
each other by Bravais translations. These bonds connect the ten Ti
ions indicated in Fig. \ref{bonds}. By the symmetry operations of
the space group $Pbnm$, the eight effective hopping matrices
between Ti ions belonging to the same $ab$ plane and the four
matrices for Ti--Ti bonds along the $c$ direction, respectively,
can be expressed by a single matrix each. For example, all
twelve hopping matrices are given by the two matrices for the
Ti--Ti bonds $mn=12$ (planar) and $mn=13$ (inter-planar),
respectively.

Despite of the different orbital ordering within the $ab$ planes and
between them,  the hopping amplitudes between the crystal-field
ground states  in and between the  planes are of the same order
of magnitude, i.\,e. roughly $|t_{12}^{00}|\approx |t_{13}^{00}|$.
In strictly cubic symmetry, those are equal (see Appendix
\ref{gshopping}). The rotation of the oxygen octahedra around the
$c$ axis, the tilting, and the distortion cause some difference
between $|t_{12}^{00}|$ and $|t_{13}^{00}|$.

For convenience, we present in Table \ref{parameters} an overview
of the parameters used in our calculation.

\subsection{The Ti--O hybridization}

Our model does not include  the covalent contribution to the
crystal field, arising from hybridization between the Ti--3$d$ and
O--2$p$ states. This mechanism mixes excited states of the static
crystal-field into the Ti$^{3+}$ ground state, i.\,e., there is an
admixture of Ti$^{2+}$ states accompanied by an admixture of holes
on the oxygen sites.

Following Ref. \onlinecite{saitoh}, we now estimate the effect of
the $pd$ hybridization. When that hybridization is absent, the
effective parameter $U_{\text{eff}}$ defines the energy difference
between the ground state of the Ti$^{3+}$ sector and the lowest
state of the Ti$^{2+}$ sector in a two-site cluster consisting of
two Ti ions. When the $pd$ hybridization is present, these two
types of $d$ states correspond to two bands, from which two $pd$
hybridized bands evolve according to the covalent crystal field.
These hybridized bands have, in general,  considerable dispersion:
Their peak-to-peak
separation, which is seen in the combined photoemission and inverse
photoemission spectra, is given by the band gap
$E_{\text{gap}}$=1.6\,eV, and the distance between the band edges
is given by the optical gap $E_{\text{opt}}$=0.2\,eV, which is
experimentally observed as the Mott gap. The mean bandwidth
between the two $pd$ hybridized bands is then
$W=E_{\text{gap}}-E_{\text{opt}}=1.4$\,eV.

Since the bands are rather dispersive, the question arises whether a
localized picture is suitable to describe, even approximately, the
LaTiO$_{3}$  system. In order to study this point, we have
analyzed the covalent crystal field of a cluster consisting of a
single Ti ion, and the six oxygen ions predominantly hybridized with
it (the calculation has been carried out for Ti No. 1 in Fig.
\ref{bonds}). This is accomplished by diagonalizing  the
Hamiltonian
\begin{eqnarray}
H_{pd}^{\nil}=
H_{\nil}^{\text{cf}}+H_{\nil}^{\text{c}}+H_{pd}^{\text{tun}},\label{Hpd}
\end{eqnarray}
for a TiO$_{6}$--cluster. Here $H^{\text{cf}}$ describes the
static crystal field, $H^{\text{c}}$
is the Coulomb interaction,
and $H_{pd}^{\text{tun}}$ is the $pd$--tunnelling,
\begin{equation}
H_{pd}^{\text{tun}}=\sum_{ni\alpha} \tilde{t}_{1n}^{i\alpha}
d_{1i}^{\dag}p_{n\alpha}^{\nil} +\text{h.\,c.}\label{pdtun}
\end{equation}
In Eq. (\ref{pdtun}), $p_{n\alpha}^{\nil}$ destroys an electron
on the $n$-th oxygen site in the $\alpha$--orbital, given in  Eq.
(\ref{pstates}). As in the calculation of the Ti--Ti hopping
amplitudes, the $pd$ hopping amplitudes,
$\tilde{t}_{1n}^{i\alpha}$, are expressed in terms of the
Slater-Koster parameters $V_{pd\sigma}$ and $V_{pd\pi}$, using the
structural data of Ref. \onlinecite{cwik}. The entire space of the
basis states of the TiO$_{6}$--cluster consists of a Ti$^{3+}$
sector where the $p$ orbitals are all occupied, and a Ti$^{2+}$
sector where there is a hole in one of the $p$ orbitals. The
eigenstates of the Hamiltonian (\ref{Hpd}) have the form
\begin{equation}
\big|\psi\big>=\sqrt{2-n_d}\,\big|d^1\big>+\sqrt{n_d-1}\,\big|d^2\big>,
\label{clusterlc}
\end{equation}
where $n_d$ is the occupation number of the Ti-$d$ shell ($1 \leq
n_d \leq 2$), $\big|d^1\big>$ is a state with a single electron in
the $d$ shell and fully occupied $p$ shells on the surrounding
oxygen ions, and $\big|d^2\big>$ is a state with two electrons in
the $d$ shell and a hole in the $p$ shell of one of the oxygen
ions. We find that in the ground state $n_d=1.334$ , i.\,e., there
is a $p$ hole on one of the neighboring oxygens with probability
of $33.4$\,\%.

This calculation allows for the analysis of the eigenstates of the
combined static and covalent crystal fields. Projecting the five
lowest eigenstates onto the Ti$^{3+}$ sector, (which corresponds to
the states $\big|d^1\big>$), gives to a very good approximation
the same eigenstates as for the static crystal field alone, see
Table \ref{covcf}. This finding explains why, despite  the
admixture of Ti$^{2+}$ states $\big|d^2\big>$, the agreement with
the experiment remains perfect, as evidenced in Eq.
(\ref{coincidence}). Indeed, the experiment measures  the
Ti$^{3+}$ part, $\big|d^1\big>$, of the combined static and
covalent crystal field, and apparently  is not sensitive to the
Ti$^{2+}$ admixture $\big|d^2\big>$. Table \ref{covcf} also shows
that the $t_{2g}$ splitting remains almost the same as in the
absence of the covalent contribution, whereas the distance between
the $t_{2g}$ and $e_g$ energies is enhanced.

Since it is extremely complicated to include in the magnetic
superexchange calculation the hopping between the $pd$ hybridized
states, we choose to consider the hopping between the Ti$^{3+}$
states only. The results listed in Table \ref{covcf}, which show
that the projections of the eigenstates of the combined static and
covalent crystal fields onto the Ti$^{3+}$ sector are almost the
same as in the static-only case, ensure that the  Ti$^{3+}$ states
we use are an appropriate starting point for the superexchange
calculation.

\subsection{Comparison with other models} \label{comparison}

It is instructive at
this point to dwell on several differences between our model and
three other calculations, reported in Refs. \onlinecite{pavarini},
\onlinecite{imada3}, and \onlinecite{eremina}.

(i) As is mentioned above, Ref.
\onlinecite{imada3} takes into account the crystal field due to
the La ions only. In addition, the intra-ionic Coulomb
correlations are approximated according to the Kanamori scheme,
which ignores the splitting of the spin-triplet states. As opposed
to this calculation, we take into account  the crystal field for
both the Ti$^{3+}$ and the Ti$^{2+}$ configurations, and employ
the full intra-ionic Coulomb correlations for the latter. The last
point is particularly important: The spin-triplet states as
intermediate Ti$^{2+}$ states cause a ferromagnetic
coupling while the spin-singlet states induce an antiferromagnetic
coupling, leading to a competition between ferromagnetic and
antiferromagnetic contributions to the magnetic exchange. On the
other hand, we omit the small $pp$ hybridization between the
oxygen states, which was included in the calculation of Ref.
\onlinecite{imada3}.

(ii) The LDA+DMFT calculation
\cite{pavarini} gives a ground state (denoted here
$\big|0''\big>$) whose projection on the experimentally-deduced
ground state, Eq. (\ref{ground2}),  is
$\big|\big<0'\big|0''\big>\big|^2=87.8\,\%$, whereas we find
$99.06\,\%$, see Eq. (\ref{coincidence}). An even larger
difference, (which is partially explained by the difference in the
ground-state orbitals) is found between the nearest-neighbor
hopping amplitudes, coupling the Ti ions in that ground state: The
values cited in Ref. \onlinecite{pavarini} are about half of the
ones we use, with the in-plane amplitude being slightly smaller
than the inter-plane one.

(iii) A recent calculation of the crystal field at room
temperature, including the covalency contribution and the
spin-orbit coupling, \cite{eremina} has yielded basically the same
$t_{2g}$ splitting scheme as ours (due to Table \ref{covcf}),
whereas the spacing between the $t_{2g}$ and $e_g$ states turned
out to be bigger than in our calculation, about 0.9\,eV.
Analogously to the way we determined the reduction of the magnetic
moment due to the spin-orbit coupling in Sec. \ref{modham}, the
ground state found in Ref. \onlinecite{eremina} has entangled
spin-up and spin-down states, i.\,e. the orbital part is not
separable from the spin part. The ground state is given there in
the fashion which has the largest possible magnetic polarization
along the quantization axis. Denoting the spin-up part of this
ground state by $\big|0'''\big>$, it turns out that the squared
overlap with the experimentally determined orbital is
$\big|\big<0'\big|0'''\big>\big|^2=92.47\,\%$. In Ref.
\onlinecite{eremina} the reduction of the G-type moment due to the
crystal field and the spin-orbit coupling is found to be 9.5\,\%,
whereas we find 14\,\%, see Sec. \ref{modham}.

We will continue the comparison with other models when we discuss results of our
calculation in Sec. \ref{clgs}.

\section{The effective spin Hamiltonian}

Our aim is to obtain from the full Hamiltonian,
$H_{mn}^{\nil}=H_{mn}^0+V_{mn}^{\nil}$, an effective spin
Hamiltonian, $h_{mn}^{\nil}$, which acts within the Hilbert space
of the four-fold degenerate ground state of the unperturbed
Hamiltonian $H_{mn}^0$.

In general, an operator which acts in the ground-state space of the two
Ti ions located at sites
$m$ and $n$, consists of linear combinations of following terms,
\begin{equation}
d_{m 0 \sigma_m'}^\dag d_{n 0 \sigma_n'}^\dag d_{n 0
\sigma_n^{\hspace*{0em}}}^{\hspace*{0em}} d_{m 0
\sigma_m^{\hspace*{0em}}}^{\hspace*{0em}},
\end{equation}
where $d^{\dagger}_{n0\sigma_{n}}$ ($d_{n0\sigma_{n}}^{\nil}$)
creates (destroys) an electron in the crystal-field ground state
at site $n$, of spin component $\sigma_{n}$. Since there is a
single electron at each Ti site, the creation and annihilation
operators can be written in terms of site spin-1/2 operators,
${\mathbf{S}}_n$,
\begin{eqnarray}
d_{n 0 \uparrow }^\dag d_{n 0
\downarrow}^{\hspace*{0em}}&=&S_n^+,\ \
d_{n 0 \downarrow }^\dag d_{n 0 \uparrow}^{\hspace*{0em}}=S_n^-, \nonumber \\
d_{n 0 \uparrow }^\dag d_{n 0 \uparrow}^{\hspace*{0em}}
&=&\mbox{$\frac{1}{2}$}+S_n^z,\ \
d_{n 0 \downarrow }^\dag d_{n 0
\downarrow}^{\hspace*{0em}}=\mbox{$\frac{1}{2}$}-S_n^z. \label{dStrafo}
\end{eqnarray}

Any operator acting within the ground-state space of the two
Ti ions  can be represented in terms of the 16 operators
\begin{eqnarray}
1&\; &(\text{constant}),\nonumber \\
S_k^\alpha&&(\text{single-ion anisotropies}),\nonumber \\
S_m^\alpha S_n^\beta&&(\text{inter-site spin couplings}),
\end{eqnarray}
where $k=m,n$ and $\alpha,\beta=x,y,z$. Since the Hamiltonian is
invariant under time-reversal,   there are no single-ion
anisotropies, and consequently the effective spin Hamiltonian
(omitting constant terms) takes the form
\begin{equation}
h_{mn}={\mathbf{S}}_m A_{mn} {\mathbf{S}}_n,
\end{equation}
where $A_{mn}^{\hspace*{0em}}\big( =A_{nm}^t\big)$ is the $3
\times 3$ superexchange matrix.   This matrix may be decomposed
into a symmetric part and an antisymmetric one. The three
components of the latter constitute the Moriya vector
${\mathbf{D}}_{mn}( =-{\mathbf{D}}_{nm}\big)$. Extracting further
the isotropic part of $A_{mn}$, i.\,e., the Heisenberg coupling
$J_{mn}$, the effective spin Hamiltonian is cast into the form
\begin{equation}
h_{mn}^{\hspace*{0em}}=J_{mn}^{\hspace*{0em}}{\mathbf{S}}_m^{\hspace*{0em}}
{\mathbf{S}}_n^{\hspace*{0em}}+{\mathbf{D}}_{mn}^{\hspace*{0em}}
\big({\mathbf{S}}_m^{\hspace*{0em}}\times{\mathbf{S}}_n^{\hspace*{0em}}\big)+
 {\mathbf{S}}_m^{\hspace*{0em}} A_{mn}^{\text{s}}
 {\mathbf{S}}_n^{\hspace*{0em}}.\label{magnetich}
\end{equation}
Here,
$A_{mn}^{\text{s}}$ represents the symmetric anisotropy. Due to
the space-group symmetries, all three types of magnetic couplings
belonging to the eight planar Ti--Ti bonds may be obtained from
those of a single bond, and so is the case for the four
inter-planar bonds, see Table \ref{hamsym}.

The various magnetic couplings appearing in Eq. (\ref{magnetich})
are obtained by perturbation theory to leading order in $V_{mn}$,
namely, to  second order in the hopping $t_{mn}$ and to first and
second order in  the spin-orbit coupling (scaled by $\lambda$). In
order to accomplish this calculation, we introduce the projection
operator $P_{mn}^0$ onto the ground-state of $H_{mn}^0$, and the
combined resolvent and projection operator $S_{mn}^{\nil}$ onto
the excited states. \cite{tak} In terms of these projection
operators, the various terms appearing in Eq. (\ref{magnetich})
acquire the following structure. The Heisenberg isotropic
exchange, to leading order in the Ti--Ti hopping, is
\begin{equation}
J_{mn}^{\hspace*{0em}}{\mathbf{S}}_m^{\hspace*{0em}}
{\mathbf{S}}_n^{\hspace*{0em}}=P_{mn}^0  H_{mn}^{\text{tun}}
 S_{mn}^{\nil} H_{mn}^{\text{tun}}  P_{mn}^0 .\label{part1}
\end{equation}

The second term in Eq. (\ref{magnetich}) is the
Dzyaloshinskii-Moriya  antisymmetric anisotropic exchange
interaction, which arises from second-order  processes in the
tunnelling Hamiltonian, and first-order processes in the
spin-orbit coupling,
\begin{eqnarray}
{\mathbf{D}}_{mn}^{\hspace*{0em}}\big({\mathbf{S}}_m^{\hspace*{0em}}
\times{\mathbf{S}}_n^{\hspace*{0em}}\big)&=&\;\;\, P_{mn}^0
H_{mn}^{\text{tun}} S_{mn}^{\nil} H_{mn}^{\text{tun}}
S_{mn}^{\nil} H_{mn}^{\text{so}}  P_{mn}^0 \nonumber\\
&& +P_{mn}^0 H_{mn}^{\text{so}} S_{mn}^{\nil} H_{mn}^{\text{tun}}
S_{mn}^{\nil} H_{mn}^{\text{tun}}  P_{mn}^0.\nonumber\\
\label{part2}
\end{eqnarray}
In fact, there are additional terms in this order, in which there
appear two Ti$^{2+}$ resolvents, e.\,g., $P_{mn}^0
H_{mn}^{\text{tun}} S_{mn}^{\nil} H_{mn}^{\text{so}} S_{mn}^{\nil}
H_{mn}^{\text{tun}}  P_{mn}^0 $. These are smaller than the ones
we keep, by an additional factor of $\simeq
\Delta_{\text{cf}}/U_{\text{eff}}=0.059$,  where
$\Delta_{\text{cf}}=0.208$\,eV is the gap between the ground state
of the single-particle crystal field and the first excited state,
see Table \ref{cf}. Following Ref. \onlinecite{shekht}, we denote
the vectors ${\mathbf{D}}_{mn}^{\hspace*{0em}}$, which refer to
the microscopic single-bond couplings of the spins, as the Moriya
vectors. The macroscopic antisymmetric anisotropic couplings
between the sublattice magnetizations of the classical ground
state (discussed in the next section)  are referred to as the
Dzyaloshinskii vectors. They are related to the Moriya vectors but
are not necessarily the same.

Finally, processes which are second-order in both the tunnelling
and the spin-orbit interaction, yield
\newpage
\end{multicols}
\begin{eqnarray}
{\mathbf{S}}_m^{\hspace*{0em}} A_{mn}^{\text{s}}
{\mathbf{S}}_n^{\hspace*{0em}}
\!+\!{\mathbf{D}}'_{mn}\big({\mathbf{S}}_m^{\hspace*{0em}}
\!\times\!{\mathbf{S}}_n^{\hspace*{0em}}\big)  &\!=\!&\;\;\,P_{mn}^0
H_{mn}^{\text{so}}  S_{mn}^{\nil} H_{mn}^{\text{tun}}
S_{mn}^{\nil} H_{mn}^{\text{tun}} S_{mn}^{\nil} H_{mn}^{\text{so}}
P_{mn}^0 \nonumber \\
&& +P_{mn}^0 H_{mn}^{\text{so}} S_{mn}^{\nil} H_{mn}^{\text{so}}
S_{mn}^{\nil} H_{mn}^{\text{tun}} S_{mn}^{\nil}
H_{mn}^{\text{tun}}  P_{mn}^0 \!+\!P_{mn}^0  H_{mn}^{\text{tun}}
S_{mn}^{\nil} H_{mn}^{\text{tun}}  S_{mn}^{\nil}
H_{mn}^{\text{so}} S_{mn}^{\nil} H_{mn}^{\text{so}}
P_{mn}^0  \nonumber \\
&&+ P_{mn}^0 H_{mn}^{\text{so}} S_{mn}^2 H_{mn}^{\text{so}}
P_{mn}^0 H_{mn}^{\text{tun}}  S_{mn}^{\nil}  H_{mn}^{\text{tun}}
P_{mn}^0\!+ \! P_{mn}^0  H_{mn}^{\text{tun}}  S_{mn}^{\nil}
H_{mn}^{\text{tun}} P_{mn}^0  H_{mn}^{\text{so}}  S_{mn}^2
H_{mn}^{\text{so}}  P_{mn}^0. \nonumber \\ \label{part3}
\end{eqnarray}
\begin{multicols}{2}
\noindent  These terms give rise to the symmetric anisotropies
$A_{mn}^{\text{s}}$, as well as to corrections
${\mathbf{D}}'_{mn}$, of order $\lambda^2$, to the Moriya vectors.
We have again omitted terms including two Ti$^{2+}$ resolvents.

As was shown in Ref. \onlinecite{shekht}, a systematic
description of the magnetic anisotropies due to the spin-orbit
interaction requires both the first and the second order processes
in $\lambda$. The technical reason being that the expectation
value of the cross product in the second term of Eq.
(\ref{magnetich}) is, in fact, also of order $\lambda$, so that
altogether the Dzyaloshinskii-Moriya interaction is at least
second order in the spin-orbit coupling. As a result, although the
antisymmetric Dzyaloshinskii-Moriya interaction alone gives rise
to spin-canting, when taken together with the symmetric
anisotropy, the system may, under specific conditions, still
preserve rotational invariance of the spins.

The detailed calculation of the various terms appearing in Eqs.
(\ref{part1}), (\ref{part2}), and (\ref{part3}) is lengthy, albeit
straight-forward. More details are given in Appendix \ref{expl}.
The values we obtain, using the parameters cited above, are listed
in Table \ref{microscres}. A comparison with spin-wave
measurements is given at the end of the following section.

\section{The classical ground state} \label{clgs}

\subsection{The magnetic order of the classical ground state}

The single-bond spin Hamiltonian, Eq. (\ref{magnetich}), is the
basis for the magnetic Hamiltonian, from which the magnetic order
of the classical ground state follows. To construct the latter,
the entire Ti-lattice is decomposed into four sublattices. Namely,
each magnetic unit cell includes four Ti ions, just as the
crystallographic unit cell. The four sublattices are hence
enumerated according to the numbers of the four Ti ions per unit
cell shown in Fig. \ref{bonds} (sublattice $i=1$ corresponds to Ti
ion 1 and its Bravais translations, etc.). Assigning a
fixed-magnitude magnetization (per site) to each sublattice,
${\mathbf{M}}_i$, one sums over all bonds which couple the four
sublattices, to obtain the {\em macroscopic} magnetic Hamiltonian
in the form
\begin{equation}
H_{\text{M}}^{\nil}=\sum_{ij}\big[I_{ij}^{\hspace*{0em}}
{\mathbf{M}}_i^{\hspace*{0em}}{\mathbf{M}}_j^{\hspace*{0em}}
+{\mathbf{D}}_{ij}^{\text{D}}\big({\mathbf{M}}_i^{\hspace*{0em}}
\times{\mathbf{M}}_j^{\hspace*{0em}}\big)+
{\mathbf{M}}_i^{\hspace*{0em}}
 \Gamma_{ij}^{\nil} {\mathbf{M}}_j^{\hspace*{0em}}\big],\label{HM}
\end{equation}
where $ij$ runs over the sublattice pairs $12,13,24,$ and $34$ of
Fig. \ref{bonds}. This summation procedure gives rise to the
macroscopic magnetic couplings: $I_{ij}^{\hspace*{0em}}$ is the
macroscopic isotropic coupling, ${\mathbf{D}}_{ij}^{\text{D}}$ are
the Dzyaloshinskii vectors (to leading order in the spin-orbit
coupling $\lambda$), which are the macroscopic antisymmetric
anisotropies, and $\Gamma_{ij}^{\nil}$ are the macroscopic
symmetric anisotropy tensors (of order $\lambda^2$). The
relations between those macroscopic couplings and the microscopic
single-bond couplings are listed in Table \ref{macrmicr}. The
inter-relations among the macroscopic couplings, which are
dictated by the symmetries of the space-group, are contained in
Table \ref{macsym}.

\begin{figure}[htb]
\leavevmode \epsfclipon \epsfxsize=7.truecm
\vbox{\epsfbox{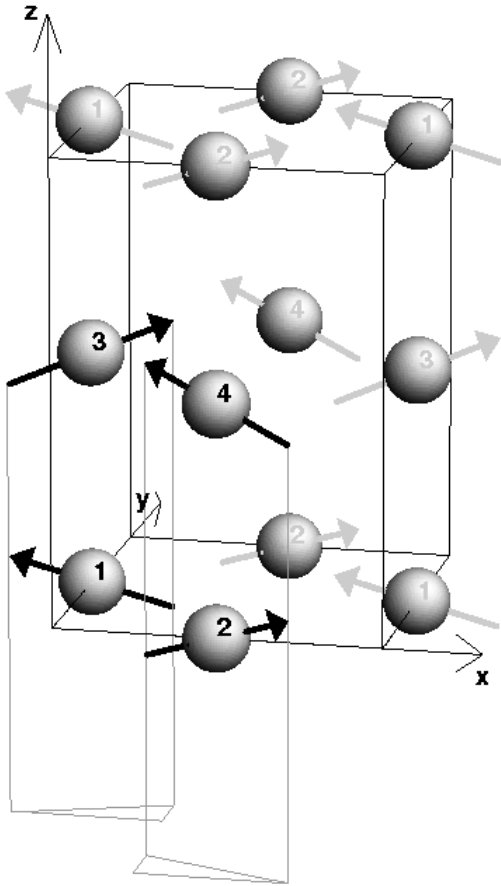}} \vspace{0.5cm}
 \caption{The magnetic order of the Ti ions in the classical
ground state of the effective spin Hamiltonian of the lattice. The
ions are enumerated according to the sublattice to which they
belong. The $x$ components of the spins order
antiferromagnetically in the G-type configuration, the $y$
components order antiferromagnetically  in the A-type one, and the
$z$ components order ferromagnetically.}\label{cgsfig}
\end{figure}

Although all four sublattice magnetizations are of equal
magnitudes, their directions are all different. Those determine
the magnetic structure of the classical ground state. In general,
according to the space group $Pbnm$ symmetries, there are four
possibilities for the symmetry of sublattice magnetizations of the
classical ground state, as listed in Table \ref{alltypes}.
\cite{meijer} In our case, the minimization of the magnetic
Hamiltonian, Eq. (\ref{HM}), yields the magnetic structure shown
in Fig. \ref{cgsfig}, drawn  according to Table \ref{cgs}. This
ground state has the following symmetry: The $x$ components of the
magnetizations order antiferromagnetically, in a G-type structure.
The $y$ components order antiferromagnetically as well, but in an
A-type structure. Finally, the $z$ components of the
magnetizations order ferromagnetically. Consequently, the
classical ground state is characterized by two canting angles,
$\varphi$ and $\vartheta$, whose values  are  given in Table
\ref{macroscres}, (see also Fig. \ref{cgsfig}). Due to the
dominating Heisenberg coupling one observes that the magnetic
structure of the classical ground state is predominantly G-type.
The easy direction along the $x$ axis and the canting angles (both
proportional to the   spin-orbit coupling $\lambda$) result from
the anisotropic couplings of the model. Those break the rotational
invariance of the magnetizations, and also cause the deviations
from the pure G-type structure.

Our magnetic structure is fully consistent with the experimental
one, as reported in Ref. \onlinecite{cwik}. This experiment (in
contrast to the one reported in Ref. \onlinecite{meijer}) reveals
that the G-type structure is indeed along the $x$ direction, while
the ferromagnetic moment is along the $z$ direction. Moreover,
since the experiment of Ref. \onlinecite{cwik} is not sensitive to
a small  moment along the $y$ axis, \cite{cwikprivat} our small
A-type antiferromagnetic order along this direction does not
contradict the data. We emphasize again that symmetry allows for
such ordering, given the G-type order along $x$ and the
ferromagnetic order along $z$. Indeed, in the YTiO$_3$-system,
which has the same space group as LaTiO$_{3}$, such order has been
detected, \cite{ulrich} but with different magnitudes of the
canting angles, which cause the ferromagnetic order to dominate.

One should note that by using naively the procedure outlined above
to obtain the energy of the  classical magnetic ground-state, one
obtains in the energy non-systematic contributions up to fourth order in the
spin-orbit coupling $\lambda$. To exemplify this point, we
consider the expectation value of $H_{\text{M}}$ for the Ti-ions
pairs $ij=12$ and $13$, expressed in terms of the angles $\varphi$
and $\vartheta$ and the superexchange  couplings,
\end{multicols}
\begin{eqnarray}
\big<H_{\text{M}}^{\nil}\big>= \big[\lambda^0\! :\big]\;&& -2
(I_{12}^{\nil}+ I_{13}^{\nil})\cos^2
\varphi \cos^2 \vartheta \nonumber\\
\big[\lambda^2\!:\big]\;&& +2(I_{12}^{\nil}-I_{13}^{\nil}) \sin^2
\varphi \cos^2 \vartheta +2(I_{12}^{\nil}+ I_{13}^{\nil}) \sin^2
\vartheta +4(D_{12}^{\text{D}\,y}+ D_{13}^{\text{D}\,y})
\cos \varphi  \cos \vartheta \sin \vartheta \nonumber \\
&&+4D_{12}^{\text{D}\,z}\cos \varphi \sin \varphi  \cos^2
\vartheta-2 (\Gamma_{12}^{xx}+\Gamma_{13}^{xx})\cos^2
\varphi \cos^2 \vartheta \nonumber\\
\big[\lambda^3\!:\big]\;&& -4D_{13}^{\text{D}\,x}\sin \varphi \cos
\vartheta \sin \vartheta  -4 \Gamma_{13}^{xy} \cos
\varphi \sin \varphi  \cos^2 \vartheta \nonumber \\
\big[\lambda^4\!:\big]\;&&  +2(\Gamma_{12}^{yy}-\Gamma_{13}^{yy})
\sin^2 \varphi \cos^2 \vartheta
+2(\Gamma_{12}^{zz}+\Gamma_{13}^{zz}) \sin^2 \vartheta  -2
\Gamma_{12}^{yz} \sin \varphi  \cos \vartheta \sin \vartheta . \label{clgsen}
\end{eqnarray}
\begin{multicols}{2}

\noindent
The leading orders of the terms are indicated in the square
brackets. The non-systematic contributions  of fourth order in
$\lambda$ are due to the couplings $\Gamma_{12}^{yy},$
$\Gamma_{13}^{yy},$ $\Gamma_{12}^{zz},$ $\Gamma_{13}^{zz},$
$\Gamma_{12}^{yz}$ (which are all of  order $\lambda^{2}$ but
 are multiplied by $\sin^{2} \vartheta$, $\sin^{2} \varphi$ and
$\sin\vartheta\sin\varphi$ which are also of order $\lambda^{2}$),
 and the
$\lambda^2$ correction of ${\mathbf{D}}_{13}^{\text{D}x}$ (which
is multiplied by $\sin\vartheta\sin\varphi$). Those contributions have
been {\em excluded} from our calculation  of the canting angles.
On the other hand, we do include  in the minimization of
$\big<H_{\text{M}}^{\nil}\big>$ terms up to the third order in
$\lambda$. This implies a systematic derivation  of the canting
angles to first order in this coupling. (The classical
ground-state energy has been found consistently, term by term, to
second order in $\lambda$. Although $\Gamma_{12}^{xx}$ and
$\Gamma_{13}^{xx}$ have been calculated only up to order
$\lambda^{2}$  and consequently, we do not have the complete
third-order term, this is of little importance when the
canting angles are determined, since those couplings appear only
with $\cos\vartheta$ and $\cos\varphi$, and therefore just cause
an energy shift in $\big<H_{\text{M}}^{\nil}\big>$.) Note also
that although the Dzyaloshinskii and Moriya vectors first appear
in linear order in $\lambda$ and the symmetric anisotropy
coefficients in quadratic order,   {\em both kinds of
anisotropies} have to be considered as they cause terms which
contribute in the same order of $\lambda$ to the classical
ground-state energy.

The values we obtain for the angles $\varphi$ and $\vartheta$ are
listed in Table \ref{macroscres}. In particular,
$\vartheta=0.80^{\circ}$, agrees, within the experimental error,
with the  values reported in Refs. \onlinecite{meijer} and
\onlinecite{fritsch}, $0.85^{\circ}$ and $1.5^{\circ}$,
respectively. (Note that $\vartheta$ is defined here with respect
to the $ab$ plane.) This canting angle causes  the weak ferromagnetic moment,
of order 0.014\,$\mu_{\text{B}}$. This value agrees with the
experimental one, within the uncertainty of the measurements which
is caused by twinning of the crystal. \cite{braden}

Recently, an attempt has been made to analyze the relation between
the anisotropy of the spin couplings and the paramagnetic
susceptibility, which also has some anisotropy. \cite{eremina} In
this work the anisotropy of the spin couplings is taken into
account by postulating phenomenologically an $xyz$ model, which
couples neighboring Ti spins and corresponds in our calculation to
the coefficients ${\mathbf{A}}_{mn}^{\text{d}}$ from Table
\ref{microscres}. A model susceptibility, which results from the
$xyz$ coupling via a molecular-field approximation (and from
single-ion as well as from covalence effects), is calculated and
then fitted with a couple of free parameters onto the measured
susceptibility. As we have shown in this section, the
antisymmetric and off-diagonal symmetric anisotropies---in the
particular case of Eq. (\ref{clgsen}), components of the
Dzyaloshinskii vectors and $\Gamma_{13}^{xy}$---can have at least
the same conceptional importance for the  magnetic properties of
LaTiO$_3$ as the $xyz$ anisotropies. Alternatively spoken, in
general the $xyz$ anisotropy is not the dominant anisotropy. This
basic argument is not restricted to the low-temperature case,
which is accompanied by the magnetic order and which we
investigate in the present paper, but refers also to the
underlying spin couplings which influence high-temperature
properties like the paramagnetic susceptibility (whereas, e.\,g.,
the difference between the structural parameters of the low- and
high-temperature case might correspond to more or less slight
differences of the spin-coupling coefficients and  of the orbital
ground state). In the way of  an extension of Ref.
\onlinecite{eremina}, the question whether it is possible to
include also  the other than $xyz$ anisotropies via free
parameters in a model susceptibility and to compare this
susceptibility to experiment, might be interesting.

\subsection{Comparison with spin-wave data} \label{spinwaves}

The magnetic order in the classical ground state is the common
starting point for a spin-wave calculation. In the case of the
spin Hamiltonian pertaining to LaTiO$_{3}$, Eq.
(\ref{magnetich}), one expects a rich spin-wave spectrum. This
calculation is currently being undertaken, and will be presented
elsewhere. \cite{unpub}
Nevertheless, our results above may be roughly compared with the
existing spin-wave data. To this end, we ignore the antisymmetric
and the symmetric anisotropies and hence assume a classical
N\'{e}el state (for which the spin-wave spectrum is gapless).

Inelastic neutron scattering has yielded the same value,
$J=15.5\,\text{meV}$, for the {\it single-bond} Heisenberg
coupling for both the Ti--Ti bonds in the $ab$ planes and those
in-between the planes. \cite{keimer} This value has been confirmed
by the evaluation of Raman spectra. \cite{iliev} Were we to average our
calculated values over the  six bonds of each Ti ion, we would
have obtained a value which is 32\,\% higher. This rather modest
discrepancy can be easily removed, by fine-tuning the model
parameters. For example, by using the value
$\Delta_{\text{eff}}=6.6$\,eV (as estimated from an LDA+DMFT
calculation  based on the recent structural data \cite{craco}), or
by using a smaller value for the Slater-Koster parameters,
$V_{pd\sigma}=-2.2$\,eV (keeping the ratio between those
 parameters fixed, $V_{pd\pi}=1.2$\,eV) instead of $-2.4$\,eV,
\cite{haverkort} or any other combined reduction of both of these
parameters. Since a detailed comparison with the data requires the
full spin-wave calculation, we do not attempt here any fine-tuning
of the model parameters.  For the purpose of the present paper, it
suffices  that the calculated Heisenberg couplings are consistent
with the experimental value, within the uncertainties of our model
parameters.

Our calculation predicts somewhat different values for the in-plane
Heisenberg coupling, $J_{12}$, and the out-of-plane one, $J_{13}$,
yielding the ratio $\delta=J_{13}/J_{12}\simeq 79\,\%$. Such an
anisotropy may be detected by  comparing with the spin-wave
dispersion, $\epsilon({\mathbf{q}})$, at selected points,
${\mathbf{q}}=(0,0,\pi)$ and ${\mathbf{q}}=(\pi,0,0)$, in the
Brillouin zone of an effective cubic lattice of a unit lattice
constant. When only this anisotropy is taken into account, then
linear spin-wave theory gives
\begin{equation}
\epsilon({\mathbf{q}})=J_{12}\sqrt{(2+\delta)^2-(\cos q_x+\cos
q_y+\delta\cos q_z)^2}.
\end{equation}
With our calculated $\delta$, we find
$\epsilon(0,0,\pi)/\epsilon(\pi,0,0)=94$\,\%, well within the
experimental error bar of about 10\,\% for the spin-wave energies,
from which the equality of the Heisenberg couplings on all bonds
has been deduced. \cite{keimerprivat} Hence, the (approximate)
isotropy of the spin-wave spectrum due to Ref.
\onlinecite{keimer}, which has been used as an argument to support
the orbital-liquid state, \cite{khaliullin} is consistent with our
model.

The calculation of Ref. \onlinecite{imada3} yielded  a
different value for the Heisenberg coupling ratio,
$\delta=106$\,\%, i.\,e. a larger coupling along the $c$ axis. This
discrepancy can be traced back to our different crystal-field
spectrum.  In our case, the hopping amplitude
between the crystal-field ground states on neighboring Ti ions is
about 10\,\% smaller for the bond 13 than for 12, see Table
\ref{effhop}. This is a geometric effect which follows from the
structural data. \cite{cwik}

From a fit to the observed spin-wave gap, of order
$\Delta=3.3\,$meV,  in conjunction with a spin model including
{\em solely} antisymmetric anisotropies, a
value of $D=1.1\,$meV has been deduced for the magnitude of the
Moriya vectors. \cite{keimer} We obtain higher
magnitudes for the Moriya vectors. However, a full spin-wave expansion based on the
Hamiltonian (\ref{magnetich})
indicates that the spin-wave gap is in fact dominated by the
symmetric anisotropies rather than by the antisymmetric ones. \cite{unpub} It
is the canting of the ordered spins with respect to each other which is
dominated by the Dzyaloshinskii vectors.

\section{Summary}

We have presented a detailed analysis of the magnetic order
pertaining to the LaTiO$_3$ system. The starting point of our
calculation is the Ti-$d$ orbital configuration which results from
the static crystal field that includes the Jahn-Teller distortion,
and which gives rise to orbital ordering as found in the
experiment. \cite{cwik} This orbital ordering rules out the
orbital-liquid picture \cite{khaliullin}  for  LaTiO$_3$, which
ignores the Jahn-Teller-like $t_{2g}$ splitting scheme and the
resulting non-degenerate orbital ground state.

Employing a perturbation expansion of this non-degenerate ground
state in the effective hopping between neighboring Ti-ions, and in
the on-site spin-orbit coup-ling, we have derived an effective spin
Hamiltonian. It includes, in addition to the Heisenberg isotropic
interaction between nearest-neighbor Ti-ions, the antisymmetric
Dzyaloshinskii-Moriya coupling, and the symmetric anisotropic
coupling. These three interactions conspire together to yield the
magnetic order. We have found, by minimizing the magnetic energy
of the classical ground state, that the magnetic order is
primarily that of a G-type antiferromagnet, with the ordered moment
along the crystallographic $a$ axis, accompanied by a weak
ferromagnetic moment along the $c$ axis. This configuration is in
 good agreement with the experimental findings. In addition, we
have found that there is a small A-type moment of the spin
components along the $b$ axis, which (although not yet detected in
experiment) is allowed by the symmetry of the system. We find
that the in-plane Heisenberg coupling energy is about 27\,\%
higher than that pertaining to the coupling between $ab$-planes. By
using these values in a spin-wave theory for the Heisenberg coup-lings, we show that
both couplings are consistent with the isotropic spin-wave
dispersion measured by inelastic neutron scattering. \cite{keimer}

Our method seems to be particularly suitable to describe the
ferromagnetic Mott insulator YTiO$_3$ as well. Preliminary calculations
(to be presented elsewhere)  indeed indicate ferromagnetic
couplings in the $ab$-planes. \cite{unpub} Since the covalent $pd$
hybridization in this system is as strong as in LaTiO$_3$,
it will be of much interest to compare the classical magnetic
Hamiltonians of the two systems, and the ensuing spin-wave
spectra.

\section{Acknowledgments}

We gratefully acknowledge discussions with M. Braden, L. Craco, M. Cwik,
M. Gr\"uninger, B. Keimer, D.\,I. Khomskii,
and L.\,H. Tjeng. This work was supported by the German-Israeli
Foundation for Research (GIF).
\end{multicols}

\appendix
\begin{multicols}{2}
\section{The Ewald summation}
\label{ewald}

The Madelung sum for the Coulomb potential in the point-charge
model is given by
\begin{equation}
V({\mathbf{r}})=-e \sum '
_{{\mathbf{l}},n}\frac{q_n}{|{\mathbf{l}}+{\mathbf{a}}_n-{\mathbf{r}}|},
\label{madelung}
\end{equation}
where ${\mathbf{r}}=(x,y,z)$ is a point on the Ti ion No. 1, whose
center is taken as the origin. In Eq. (\ref{madelung}),
$\mathbf{l}$ are the Bravais translations, $\mathbf{a}_n$ are the
basis vectors of the unit cell, and $q_n$ are the corresponding
point charges. The prime on the sum symbol indicates that the ion
at the origin, $\mathbf{l}=\mathbf{a}_n=\mathbf{0}$, is omitted.
This sum converges very slowly. One therefore uses the Ewald
summation, \cite{ewald} (see also  Ref. \onlinecite{ziman}) where
the sum (\ref{madelung}) is mapped onto two sums which
converge much better, and which can be  computed to high accuracy. Using the
Ewald summation, the Madelung potential can be expressed in the
form
\end{multicols}
\begin{eqnarray}
V({\mathbf{r}})=-\,e \sum_{{\mathbf{g}}\neq {\mathbf{0}}} \frac{4
\pi}{V_{\text{e}}g^2}\,e^{-\frac{g^2}{4G^2}+ i
{\mathbf{gr}}}\sum_n q_n e^{-i{\mathbf{ga}}_n }
-e
\sum_{{\mathbf{l}},n}'\frac{q_n}{|{\mathbf{l}}+{\mathbf{a}}_n-
{\mathbf{r}}|}\,\text{erfc}\,(G|{\mathbf{l}}+{\mathbf{a}}_n-{\mathbf{r}}|)
+e \frac{q_1}{r}\, \text{erf}\,(Gr).\label{pot}
\end{eqnarray}
\begin{multicols}{2}
\noindent  Here ${\mathbf{g}}$ are the vectors of the reciprocal
Bravais lattice, whose basis vectors are $(2\pi/a,0,0)$,
$(0,2\pi/b,0)$, and $(0,0,2\pi/c)$, $V_{\text{e}}=abc$ is the
volume of the unit cell, and $G$ is a frequency cutoff. The value
of $V({\mathbf{r}})$ is of course independent of $G$. This cutoff
is chosen such that the sum over the real-space lattice and the
one over the reciprocal lattice can be stopped after about the
same number of sites, when the required numerical precision is
reached. In Eq. (\ref{pot}), erf and erfc are  the error functions
\begin{equation}
\text{erf}\,(z)=1-\text{erfc}\,(z)=\frac{2}{\sqrt{\pi}}\int_0^z e^{-t^2} dt.
\end{equation}
The Ewald construction requires the neutrality condition
\begin{equation}
\sum_n q_n=0,
\end{equation}
which   is fulfilled in our case.

In order to find  the spectrum and the eigenstates of the static
crystal field, we have replaced the potential $V({\mathbf {r}})$
by the pseudo-potential $V_{\text{ps}}({\mathbf{r}})$, which is
its Taylor expansion including the second and fourth orders in
$\mathbf{r}$. These are the Taylor orders which have non-trivial
matrix elements with respect to the $d$ orbitals. \cite{newman}
For instance, the second (fourth) Taylor order includes also terms
which are proportional to $xy$ ($x^3y,x^2y^2$). For the Taylor
expansion we use ($r=|{\mathbf{r}}|$)
\begin{equation}
\frac{1}{r}\,\text{erf}\,(Gr)=\frac{2G}{\sqrt{\pi}} \left[1-
\frac{(Gr)^2}{3}+\frac{(Gr)^4}{10}+... \right].
\end{equation}
The potential  $V_{\text{ps}}({\mathbf{r}})$ is a harmonic
function, invariant under inversion of the coordinates. The
diagonalization of the matrix
$\big<\gamma\big|V_{\text{ps}}({\mathbf{r}})\big|\gamma'\big>$,
where $\gamma$ and $\gamma'$ denote the orthorhombic $d$ orbitals, gives the
results listed in Table \ref{cf} for the static crystal field.
This calculation requires the second and fourth moments of the
effective ionic radius, defined by
\begin{equation}
\int f^2(r) r^{2+n} dr=\big<r^n\big>, \quad n=2,4,
\end{equation}
where $f(r)$ denotes the radial part of the $d$-orbitals.
\cite{norm}

\section{The hopping amplitudes between the crystal-field ground states}
\label{gshopping}

As is mentioned in the text, the effective Ti--Ti hopping matrix
elements between the crystal-field ground states in the $ab$
planes are of the same order of magnitude as those between
planes, i.e., $|t_{12}^{00}|\approx |t_{13}^{00}|$. This is a
somewhat surprising result in view of the fact that there is
ferro-orbital order in the planes and (canted) antiferro-orbital
order between them. However, as we show here, in strictly cubic
symmetry one has $|t_{12}^{00}|= |t_{13}^{00}|$. The deviations
from cubic symmetry cause the slight difference between these two
hopping amplitudes.

Let us hence consider the cubic case, and employ the coordinate
system $x',y',z$ in which the Ti sites 1 and 2 are on the $x'$
axis (see Fig. \ref{bonds}; these coordinates are rotated by
45$^\circ$ compared to the orthorhombic ones). The crystal-field
ground states are now linear combinations of the three degenerate
$t_{2g}$ orbitals
\begin{equation}
\big|y'z\big>,\big|x'z\big>,\big|x'y'\big>.
\end{equation}

\begin{figure}
\begin{center}
\leavevmode \epsfclipon \epsfxsize=7.truecm
\vbox{\epsfbox{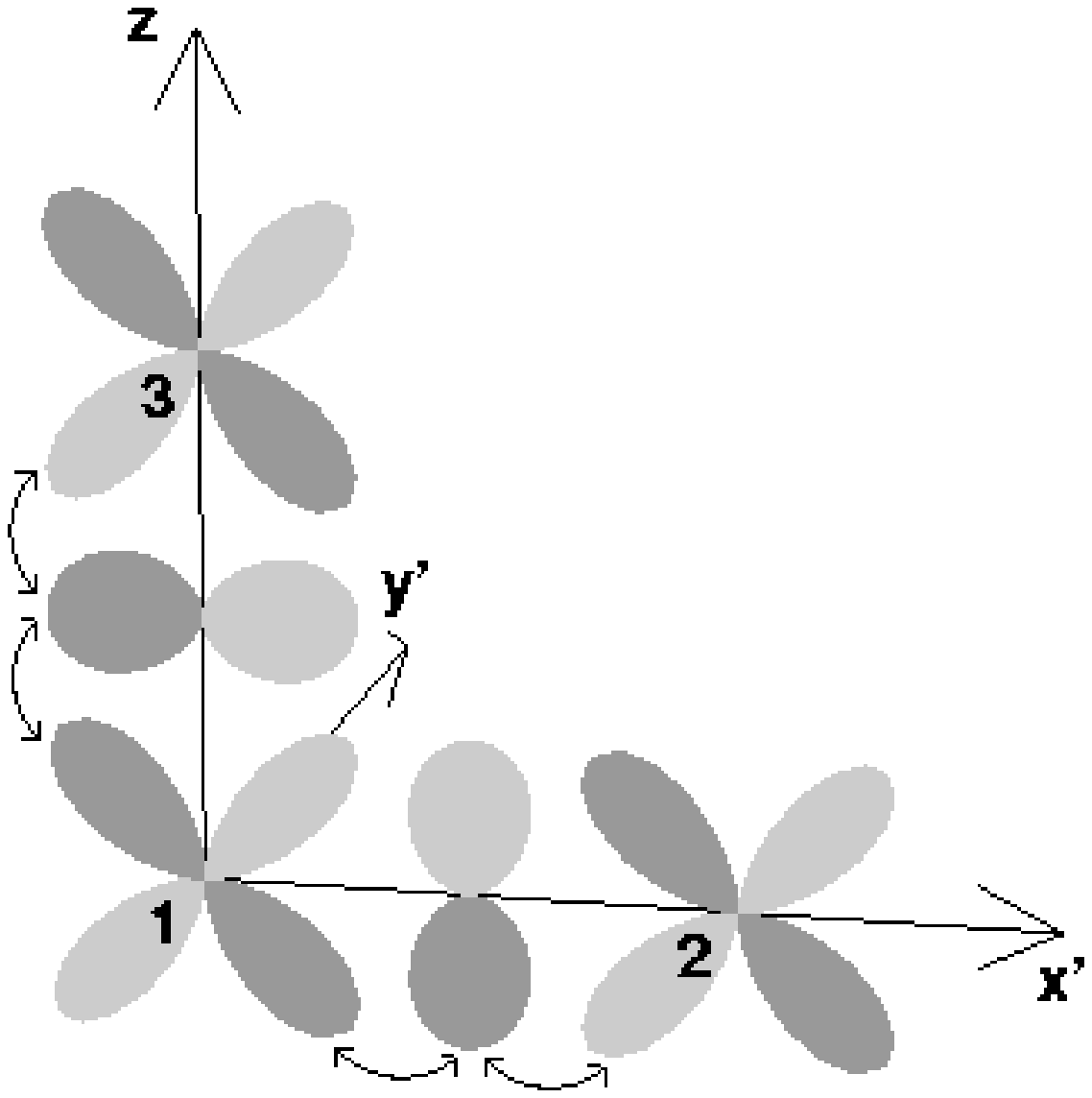}}
\end{center}
\caption{In cubic symmetry, there is one  $t_{2g}$ orbital per
each Ti--O--Ti bond, which cannot participate in the hopping
between the Ti and the O ions because of the parity of the O-2$p$
orbitals. This is shown in panels (a)--(c), for each of the three
$t_{2g}$ orbitals, respectively. (a) The hopping between the
orbitals $\big|y'z\big>$ at the Ti sites 1 and 2 is not possible,
since on the intermediate oxygen site there is no $p$ orbital with
a parity that would allow such hopping. For the same reason, an
electron cannot hop from the $p$ orbital $\big|x'\big>$, which is
shown between the Ti sites 1 and 2, to any of the $t_{2g}$
orbitals of the Ti sites. On the other hand, the hopping between
the orbitals $\big|y'z\big>$ at the Ti sites 1 and 3, which is
mediated by the orbital $\big|y'\big>$ on the intermediate oxygen
site, is possible. (b) Analogously to the case (a), the hopping
between the orbitals $\big|x'y'\big>$ at the Ti sites 1 and 3 is
not permitted, because on the intermediate oxygen site there is no
appropriate $p$ orbital. Likewise, an electron cannot hop from the
$p$ orbital $\big|z\big>$, which is shown between the Ti sites 1
and 3, to any of the $t_{2g}$ orbitals of the Ti sites. The
(allowed) hopping between the orbitals $\big|x'y'\big>$ at the Ti
sites 1 and 2 is mediated by the orbital $\big|y'\big>$ on the
intermediate oxygen site. (c) The hopping between the orbitals
$\big|x'z\big>$ at the Ti sites 1 and 2 is mediated by the orbital
$\big|z\big>$ on the intermediate oxygen site. The hopping between
the orbitals $\big|x'z\big>$ at the Ti sites 1 and 3 is mediated
by the orbital $\big|x'\big>$ on the intermediate oxygen site.}
\label{t2gfigs}
\end{figure}

The hopping amplitudes are proportional to the overlap of the two
pertaining orbitals. Let us consider for simplicity the
base-orbitals according to Eq. (\ref{ground2}), i.e.,
\begin{eqnarray}
\big|1\big>=\big|2\big>&\!=\!&\mbox{$
\frac{1}{\sqrt{3}}$}\big(\quad \big|y'z \big>-\big| x'z\big>
+\big|x'y' \big> \big), \nonumber \\
\big|3\big>&\!=\!&\mbox{$\frac{1}{\sqrt{3}}$}\big(\!-\big|y'z
\big>+\big| x'z\big>+\big|x'y' \big> \big),
\end{eqnarray}
where 1, 2 and 3 denote the relevant Ti--ions (see Fig.
\ref{bonds}). Were we to find the direct hopping between the Ti
sites, we would have obtained for the overlaps the result
\begin{equation}
\big<1\big|2\big>=1, \quad
\big<1\big|3\big>=-\mbox{$\frac{1}{3}$},
\end{equation}
which reflects the ferro-orbital and nearly antiferro-orbital
order, respectively. However, the {\em effective} Ti--Ti hopping
that we consider is mediated by the oxygens located between the Ti
ions. Then, in strictly cubic symmetry,  for each pair of Ti ions,
one of the three $t_{2g}$ orbitals cannot hybridize. This
`inactiveness'  of one of the orbitals \cite{khaliullin} is a
direct consequence of the cubic symmetry, as is portrayed in Fig.
\ref{t2gfigs}, and is the source
of peculiar hidden symmetries in the cubic Hamiltonian.
\cite{harris03,harris04} In our example,  the orbital
$\big|y'z\big>$ is inactive for the 12--bond, while for the bond
13 the inactive orbital is $\big|x'y'\big>$, and consequently
\begin{eqnarray}
\big<1\big|2\big>&\to &\mbox{$\frac{1}{3}$}\big(-\big< x'z\big|+
\big<x'y' \big|  \big)\big(-\big| x'z\big>+\big|x'y' \big>  \big)
 = \quad \! \! \mbox{$\frac{2}{3}$}, \nonumber \\
\big<1\big|3\big>&\to &\mbox{$\frac{1}{3}$}\big(\quad \,\,\big<
y'z\big|- \big< x'z\,\big| \big)\big(-\big| y'z\big>+ \big|
x'z\,\big> \big)\, = -
\mbox{$\frac{2}{3}$}, \nonumber \\
\end{eqnarray}
leading to $|t_{12}^{00}|= |t_{13}^{00}|$.

\section{The explicit calculation of the exchange couplings} \label{expl}

Here we document the technical details of the perturbation
calculation, that yields the effective spin Hamiltonian.

As explained in Sec. II, we consider a cluster of two
nearest-neighbor Ti ions. The Hamiltonian of this cluster, given
in Eqs. (\ref{unperturbed}) and (\ref{perturbed}), is expressed in
terms of the operators $d^{\dagger}_{ki\sigma}$ ($d_{ki\sigma}^{\nil})$
which create (destroy) an electron  in the crystal-field
eigen-orbital $i$ with spin component $\sigma$, on the Ti ion
located at site $k$. However, it is more convenient to treat the
two-electron states (which appear in the intermediate stages of
the perturbation expansion) using the orthorhombic basis, Eq.
(\ref{dbasis}). We denote the operators pertaining to this basis
by $c^{\dagger}_{k\gamma\sigma}$ ($c_{k\gamma\sigma}^{\nil}$), where
$\gamma $ enumerates the orthorhombic orbitals. The first part of
this Appendix is devoted to the transformation of the Hamiltonian
between the two schemes, and the diagonalization of the
two-electron states. In the second part, we summarize the detailed
expressions of the various terms resulting from the perturbation
expansion.

\subsection{The Hamiltonian}

Denoting the  matrix of the crystal-field Hamiltonian in the
orthorhombic basis   by $V(k)$,  we have
\begin{eqnarray}
H^{\text{cf}}= \sum_{k \gamma_1 \gamma_2 \sigma} \hspace*{-0.5em}
V_{\gamma_1 \gamma_2}^{\nil}(k) c^{\dagger}_{k \gamma_1
\sigma}c_{k\gamma _2\sigma}^{\nil}.
\end{eqnarray}
The matrices $V(k)$ are real and symmetric. We next introduce the
(unitary and real) matrix $W(k)$ which diagonalizes the
crystal-field Hamiltonian, bringing it to the form
\begin{eqnarray}
H^{\text{cf}}=\sum_{k i \sigma} E_i^{\nil} d^{\dagger}_{k i
\sigma}d_{ki\sigma}^{\nil},
\end{eqnarray}
where $E_{i}$ are the crystal-field eigenvalues, listed in Table
\ref{cf}.  These single-particle energies are shifted so that
$E_0=0\,\text{eV}$, $E_1=0.209\,\text{eV}$, etc. The relations
between the operators $d^{\dagger}_{ki \sigma}$ and
$c^{\dagger}_{k\gamma\sigma}$ are hence given by
\begin{eqnarray}
d_{ki\sigma}^{\dag} =\sum_{\gamma} W_{i\gamma}^{\nil}(k)
c_{k\gamma\sigma}^{\dag}, \quad c_{k\gamma\sigma}^{\dag}=\sum_{i}
W_{\gamma i}^{t}(k) d_{ki\sigma}^{\dag}, \label{CFDIAG}
\end{eqnarray}
such that
\begin{eqnarray}
W(k)V(k)W^t(k)=E , \label{cftrafo}
\end{eqnarray}
with $E=\text{diag}\,E_i$. The diagonalizing matrix pertaining to
site 1,  $W(1)$, is given in Table \ref{cf}. All other $W(k)$
and $V(k)$ follow from the symmetry properties of the unit cell, and are
given by

\newpage
\begin{center}
(a) The hopping between the orbitals $\big|y'z\big>$\\[2ex]
\end{center}
\begin{figure*}
\begin{center}
\leavevmode \epsfclipon \epsfxsize=7.truecm \vbox{\epsfbox{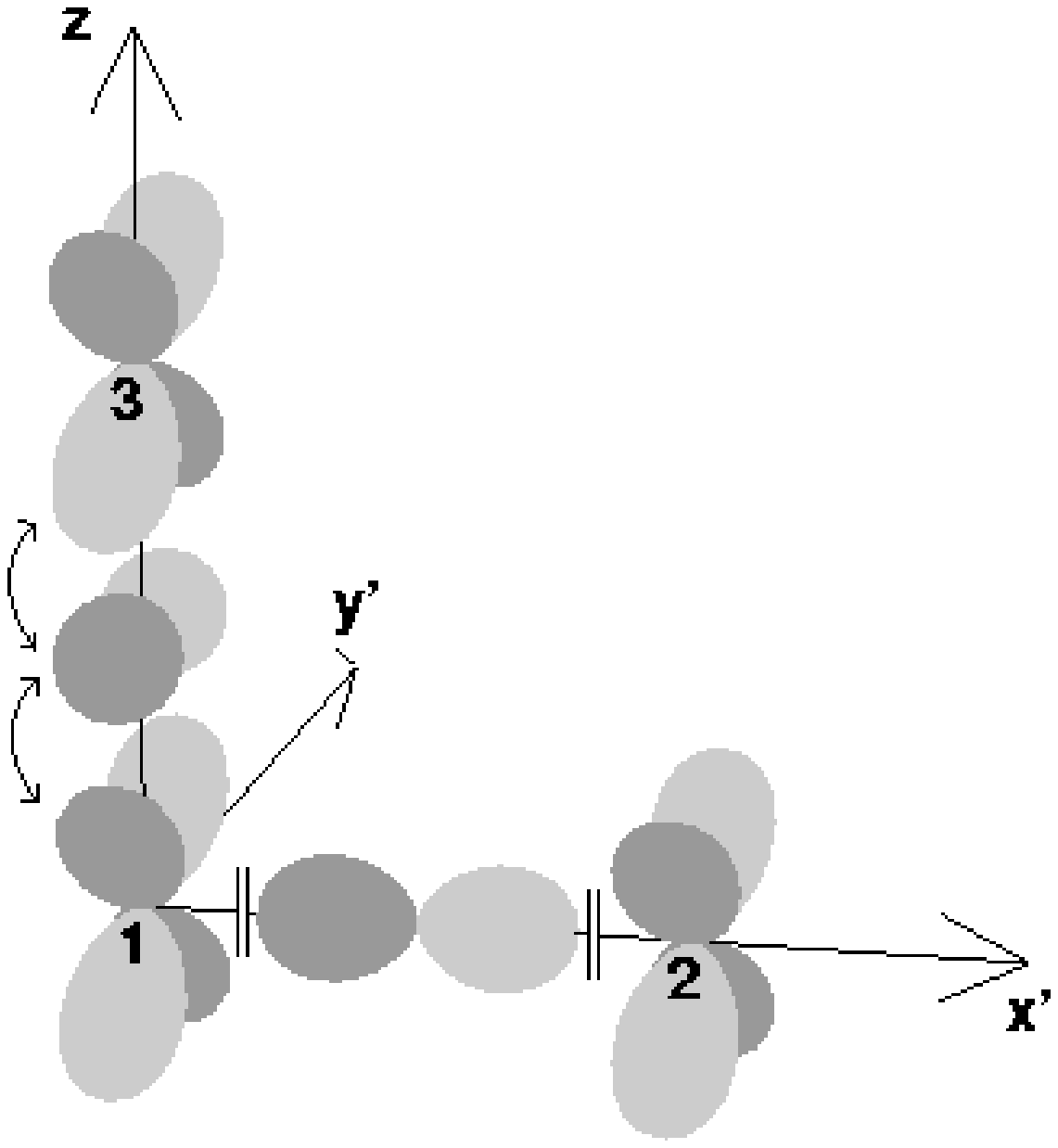}}
\vspace*{0.5cm}
\end{center}
\end{figure*}

\begin{center}
(b) The hopping between the orbitals $\big|x'y'\big>$\\[2ex]
\end{center}
\begin{figure*}
\begin{center}
\leavevmode \epsfclipon \epsfxsize=7.truecm \vbox{\epsfbox{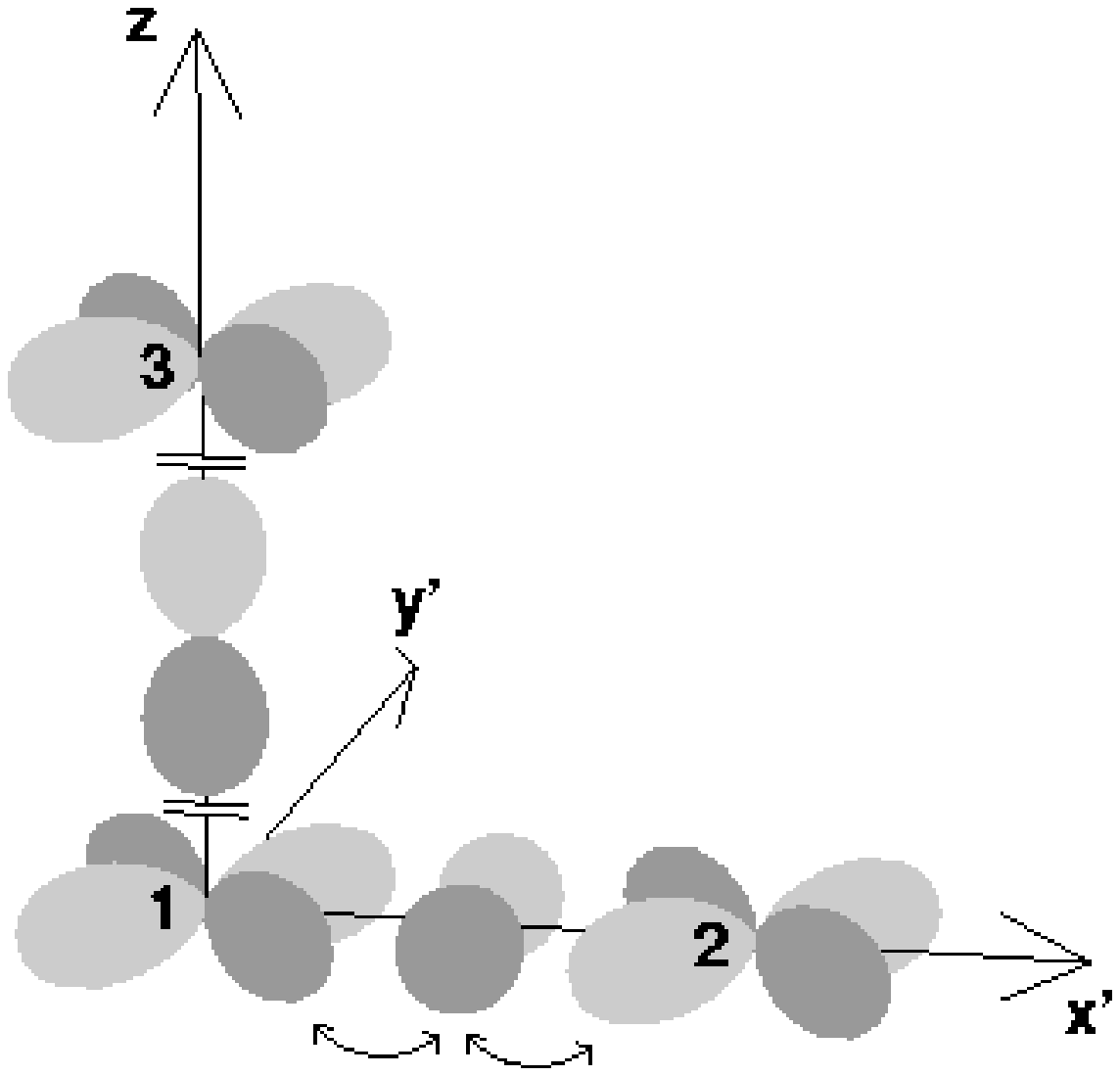}}
\vspace*{0.5cm}
\end{center}
\end{figure*}

\vspace*{4cm}
\begin{center}
(c) The hopping between the orbitals $\big|x'z\big>$\\[2ex]
\end{center}

\newpage
\begin{eqnarray}
W(2)&=&\left[\begin{array}{ccccc}
-&+&+&-&+\\
-&+&+&-&+\\
-&+&+&-&+\\
-&+&+&-&+\\
-&+&+&-&+
\end{array} \right]\otimes W(1),\nonumber\\
V(2)&=&\left[\begin{array}{ccccc}
+&-&-&+&-\\
-&+&+&-&+\\
-&+&+&-&+\\
+&-&-&+&-\\
-&+&+&-&+
\end{array} \right]\otimes V(1),\nonumber\\
W(3)&=&\left[\begin{array}{ccccc}
+&+&-&-&+\\
+&+&-&-&+\\
+&+&-&-&+\\
+&+&-&-&+\\
+&+&-&-&+
\end{array} \right]\otimes W(1),\nonumber\\
V(3)&=&\left[\begin{array}{ccccc}
+&+&-&-&+\\
+&+&-&-&+\\
-&-&+&+&-\\
-&-&+&+&-\\
+&+&-&-&+
\end{array} \right]\otimes V(1),
\end{eqnarray}
where we have used the direct matrix product $a=b \otimes c$,
i.e.,  $a_{ij}=b_{ij}c_{ij}$.

\subsubsection{The Coulomb Hamiltonian}

\noindent The Coulomb Hamiltonian, in the orthorhombic basis, is
given by
\begin{eqnarray}
H^{\text{c}}=\frac{1}{2}\hspace*{-0.5em} \sum_{k
\sigma_{1}\sigma_{2} \atop
\gamma_{1}\gamma_{2}\gamma_{3}\gamma_{4}}
\hspace*{-0.5em}U_{\gamma_{1}\gamma_{2}\gamma_{3}\gamma_{4}}^{\nil}
c^{\dagger}_{k\gamma_{1}\sigma_{1}}
c^{\dagger}_{k\gamma_{2}\sigma_{2}}c_{k\gamma_{3}\sigma_{2}}^{\nil}
c_{k\gamma_{4}\sigma_{1}}^{\nil}.
\end{eqnarray}
In order to specify its matrix elements in the 3$d^2$ sector (taken  from Ref.
\onlinecite{slater}) we construct the triplet
($\Psi_{T}^{\gamma\gamma '}$) and the singlet
($\Psi_{S}^{\gamma\gamma '}$) two-particle states in the
orthorhombic basis,
\begin{eqnarray}
\Psi_{T}^{\gamma\gamma '}(k;\sigma\sigma
')&=&\sqrt{\frac{1}{2}}\Bigl
(c^{\dagger}_{k\gamma\sigma}c^{\dagger}_{k\gamma '\sigma
'}+c^{\dagger}_{k\gamma\sigma '}c^{\dagger}_{k\gamma '\sigma
}\Bigr )\nonumber\\
&=&-\Psi_{T}^{\gamma '\gamma }(k;\sigma\sigma ')
,\nonumber\\
\Psi_{S}^{\gamma\gamma '}(k;\sigma\sigma
')&=&\sqrt{\frac{1}{2}}\Bigl
(c^{\dagger}_{k\gamma\sigma}c^{\dagger}_{k\gamma '\sigma
'}-c^{\dagger}_{k\gamma\sigma '}c^{\dagger}_{k\gamma '\sigma
}\Bigr
)\nonumber\\
&=&\Psi_{S}^{\gamma '\gamma }(k;\sigma\sigma '),\nonumber\\
\Psi_{S}^{\gamma\gamma}(k;\sigma\sigma
')&=&c^{\dagger}_{k\gamma\sigma}c^{\dagger}_{k\gamma\sigma
'}.\label{trisin}
\end{eqnarray}
Altogether,  there are 10 triplets and 15 singlets (in the
$\sigma'=-\sigma$ sector). Enumerating the triplet states in the
following order,
\end{multicols}
\begin{eqnarray}
|1\rangle &=&\Psi_{T}^{52},\ |2\rangle
=\Psi_{T}^{53},\ |3\rangle =\Psi_{T}^{41},\
|4\rangle =\Psi_{T}^{23},\ |5\rangle
=\Psi_{T}^{54} ,\nonumber\\
|6\rangle &=&\Psi_{T}^{13},\ |7\rangle
=\Psi_{T}^{24}, \ |8\rangle =\Psi_{T}^{21},\
|9\rangle = \Psi_{T}^{34},\ |10\rangle
=\Psi_{T}^{51},\label{notationtriplet}
\end{eqnarray}
the Coulomb Hamiltonian in the triplet sector becomes
\begin{eqnarray}
U_T^{\nil}=
\left[\begin{array}{cccccccccc}
A-8B&0&0&0&0&0&0&0&0&0\\
0&A-5B&3B&3B\sqrt{3}&0&0&0&0&0&0\\
0&3B&A-5B&3B\sqrt{3}&0&0&0&0&0&0\\
0&3B\sqrt{3}&3B\sqrt{3}&A+B&0&0&0&0&0&0\\
0&0&0&0&A-5B&3B&-3B\sqrt{3}&0&0&0\\
0&0&0&0&3B&A-5B&-3B\sqrt{3}&0&0&0\\
0&0&0&0&-3B\sqrt{3}&-3B\sqrt{3}&A+B&0&0&0\\
0&0&0&0&0&0&0&A-8B&0&0\\
0&0&0&0&0&0&0&0&A-5B&-6B\\
0&0&0&0&0&0&0&0&-6B&A+4B
\end{array}\right]=U_T^t.
\end{eqnarray}
Here, $A$, $B$, and $C$ are the Racah parameters, given by
combinations of  the   Slater integrals $F_2$ and $F_4$,
\begin{eqnarray}
A=A_0-\frac{49}{441}F_4, \quad
B=\frac{1}{49}F_2-\frac{5}{441}F_4, \quad
C=\frac{35}{441}F_4.
\end{eqnarray}
The parameter  $A_0$ is determined such that upon diagonalizing
simultaneously  the Coulomb Hamiltonian and the crystal-field one,
the lowest state has the energy $U_{\text{eff}}$ as explained in
Sec. \ref{modham}. Enumerating the singlets in the order
\begin{eqnarray}
|1\rangle &=&\Psi_{S}^{55},\ |2\rangle =\Psi_{S}^{22},\ |3\rangle
=\Psi_{S}^{33},|4\rangle
=\Psi_{S}^{44},\ |5\rangle =\Psi_{S}^{11},\nonumber\\
|6\rangle &=&\Psi_{S}^{52},\ |7\rangle =\Psi_{S}^{51},\ |8\rangle
=\Psi_{S}^{34},\ |9\rangle =\Psi_{S}^{21},\ |10\rangle
=\Psi_{S}^{41},\nonumber\\
|11\rangle &=&\Psi_{S}^{53},\ |12\rangle =\Psi_{S}^{23}, \
|13\rangle =\Psi_{S}^{13},\ |14\rangle =\Psi_{S}^{54}, \
|15\rangle =\Psi_{S}^{24},\label{notationsinglet}
\end{eqnarray}
the Coulomb Hamiltonian in the singlet sector becomes
\begin{eqnarray}
U_S^{\nil}=\left[\begin{array}{ccc} U_{S1}&0&0\\
0&U_{S2}&0\\
0&0&U_{S3}
\end{array}\right]=U_S^t,
\end{eqnarray}
where
\begin{eqnarray}
U_{S1}=\left[\begin{array}{cccccc} A+4B+3C&4B+C&3B+C&3B+C&C&0\\
4B+C&A+4B+3C&B+C&B+C&4B+C&0\\3B+C&B+C&A+4B+3C&3B+C&3B+C&-B\sqrt{6}\\
3B+C&B+C&3B+C&A+4B+3C&3B+C&B\sqrt{6}\\
C&4B+C&3B+C&3B+C&A+4B+3C&0\\0&0&-B\sqrt{6}&B\sqrt{6}&0&A+2C
\end{array}\right], \nonumber
\end{eqnarray}
\begin{eqnarray}
U_{S2}=\left[\begin{array}{ccc}
A+4B+2C&0&0\\0&A+B+2C&2B\sqrt{3}\\0&2B\sqrt{3}&A+2C
\end{array}\right],\nonumber
\end{eqnarray}
\begin{eqnarray}
U_{S3}=\left[\begin{array}{cccccc}
A+B+2C&3B&-B\sqrt{3}&0&0&0\\
3B&A+B+2C&B\sqrt{3}&0&0&0\\
-B\sqrt{3}&B\sqrt{3}&A+3B+2C&0&0&0\\
0&0&0&A+B+2C&-3B&-B\sqrt{3}\\0&0&0&-3B&A+B+2C&-B\sqrt{3}\\
0&0&0&-B\sqrt{3}&-B\sqrt{3}&A+3B+2C
\end{array}\right].
\end{eqnarray}
\begin{multicols}{2}

\subsubsection{The spin-orbit Hamiltonian}
\noindent Written in the orthorhombic basis, the spin-orbit
Hamiltonian is
\begin{eqnarray}
H^{\text{so}}=\frac{\lambda}{2}\sum_{\alpha\gamma\gamma ' \atop k
\sigma\sigma'}L_{\gamma\gamma
'}^{\alpha}\sigma^{\alpha}_{\sigma\sigma
'}c_{k\gamma\sigma}^{\dag}c_{k\gamma '\sigma '}^{\nil},
\end{eqnarray}
where $\alpha$ takes the values $x, y$, and $z$, $\sigma^{\alpha}$
are the Pauli matrices, and the angular momentum matrices
$L^{\alpha}_{\gamma\gamma '}$  are
\begin{eqnarray}
L^x&=&\left[\begin{array}{cccccc}
0&0&0&-i&0\\
0&0&i\sqrt{3}&0&0\\
0&-i\sqrt{3}&0&0&-i\\
i&0&0&0&0\\
0&0&i&0&0
\end{array}\right], \nonumber \\
L^y&=&\left[\begin{array}{cccccc}
0&0&i&0&0\\
0&0&0&-i\sqrt{3}&0\\
-i&0&0&0&0\\
0&i\sqrt{3}&0&0&-i\\
0&0&0&i&0
\end{array}\right],\nonumber \\
L^z&=&\left[\begin{array}{cccccc}
0&0&0&0&2i\\
0&0&0&0&0\\
0&0&0&i&0\\
0&0&-i&0&0\\
-2i&0&0&0&0
\end{array}\right].
\end{eqnarray}
Transformed into the crystal-field eigenstates, the spin-orbit
Hamiltonian takes the form
\begin{eqnarray}
H^{\text{so}}=\frac{\lambda}{2}\sum_{\alpha ii' \atop
k\sigma\sigma'}L^{\alpha}_{ii'}(k)\sigma^{\alpha}_{\sigma\sigma
'}d^{\dagger}_{ki\sigma}d^{\nil}_{ki'\sigma '},
\end{eqnarray}
with
\begin{eqnarray}
L^{\alpha}_{ii'}(k)=\sum_{\gamma\gamma
'}W^{\nil}_{i\gamma}(k)L^{\alpha}_{\gamma\gamma '}W^t_{\gamma 'i'}(k).
\end{eqnarray}
Note that the relation $L^{\alpha}_{\gamma\gamma
'}=-L^{\alpha}_{\gamma '\gamma}$ implies that
$L^{\alpha}_{ii'}(k)=-L^{\alpha}_{i'i}(k).$

\subsubsection{The diagonalization of the two-electron states}

\noindent When there are two electrons on the same Ti ion (at site
$k$), their state is described by
$d^{\dagger}_{ki\sigma}d^{\dagger}_{kj\sigma '}$. Using Eqs.
(\ref{CFDIAG}) and (\ref{trisin}), we re-write this state in terms
of the singlet and triplet states,
\begin{eqnarray}
&&d^{\dagger}_{ki\sigma}d^{\dagger}_{kj\sigma
'}=c^{\dagger}_{k\gamma\sigma}c^{\dagger}_{k\gamma '\sigma
'}W_{i\gamma}^{\nil}(k)W_{j\gamma '}^{\nil}(k)\nonumber\\
&=&W_{i\gamma }^{\nil}(k)W_{j\gamma '}^{\nil}(k)\Bigl
\{\Psi_{S}^{\gamma\gamma}(k;\sigma\sigma ')\delta_{\gamma\gamma
'}^{\nil}\nonumber\\
&&+\sqrt{\frac{1}{2}}\Bigl [\Psi_{T}^{\gamma\gamma
'}(k;\sigma\sigma ')+\Psi_{S}^{\gamma\gamma '}(k;\sigma\sigma
')\Bigr ](1-\delta_{\gamma\gamma '}^{\nil})\Bigr \},\label{cc}
\end{eqnarray}
where we have omitted for brevity the summation notations.
Adopting the enumeration conventions Eqs. (\ref{notationtriplet})
and (\ref{notationsinglet}), this state can be cast conveniently
into the form
\begin{eqnarray}
d^{\dagger}_{ki\sigma}d^{\dagger}_{kj\sigma '} &=&\sum_{\mu
=1}^{10}w_{T}^{\mu}(k;ij)\Psi_{T}^{\mu}(k;\sigma\sigma
')\nonumber\\
&&+\sum_{\mu
=1}^{15}w_{S}^{\mu}(k;ij)\Psi_{S}^{\mu}(k;\sigma\sigma ').
\end{eqnarray}
Here we have introduced  the 10-dimensional vector $w_{T}$, whose
components are
\begin{eqnarray}
w_{T}(k;ij)=\sqrt{\frac{1}{2}}\Bigl
[&&W_{i5}(k)W_{j2}(k)-W_{i2}(k)W_{j5}(k),\nonumber\\
&&W_{i5}(k)W_{j3}(k)-W_{i3}(k)W_{j5}(k),\nonumber\\
&&W_{i4}(k)W_{j1}(k)-W_{i1}(k)W_{j4}(k),\nonumber\\
&&W_{i2}(k)W_{j3}(k)-W_{i3}(k)W_{j2}(k),\nonumber\\
&&W_{i5}(k)W_{j4}(k)-W_{i4}(k)W_{j5}(k),\nonumber\\
&&W_{i1}(k)W_{j3}(k)-W_{i3}(k)W_{j1}(k),\nonumber\\
&&W_{i2}(k)W_{j4}(k)-W_{i4}(k)W_{j2}(k),\nonumber\\
&&W_{i2}(k)W_{j1}(k)-W_{i1}(k)W_{j2}(k),\nonumber\\
&&W_{i3}(k)W_{j4}(k)-W_{i4}(k)W_{j3}(k),\nonumber\\
&&W_{i5}(k)W_{j1}(k)-W_{i1}(k)W_{j5}(k)\Bigr ]
\end{eqnarray}
and the 15-dimensional vector $w_{S}$,
\begin{eqnarray}
w_{S}(k;ij)=\Bigl
[&&W_{i5}(k)W_{j5}(k),W_{i2}(k)W_{j2}(k),\nonumber\\
&&W_{i3}(k)W_{j3}(k),W_{i4}(k)W_{j4}(k),\nonumber\\
&&W_{i1}(k)W_{j1}(k)\Bigr ],\
\end{eqnarray}
for the entries $\mu =1,...,5,$  and
\begin{eqnarray}
w_{S}(k;ij)=\sqrt{\frac{1}{2}}\Bigl
[&&W_{i5}(k)W_{j2}(k)+W_{i2}(k)W_{j5}(k),\nonumber\\
&&W_{i5}(k)W_{j1}(k)+W_{i1}(k)W_{j5}(k),\nonumber\\
&&W_{i3}(k)W_{j4}(k)+W_{i4}(k)W_{j3}(k),\nonumber\\
&&W_{i2}(k)W_{j1}(k)+W_{i1}(k)W_{j2}(k),\nonumber\\
&&W_{i4}(k)W_{j1}(k)+W_{i1}(k)W_{j4}(k),\nonumber\\
&&W_{i5}(k)W_{j3}(k)+W_{i3}(k)W_{j5}(k),\nonumber\\
&&W_{i2}(k)W_{j3}(k)+W_{i3}(k)W_{j2}(k),\nonumber\\
&&W_{i3}(k)W_{j1}(k)+W_{i1}(k)W_{j3}(k),\nonumber\\
&&W_{i5}(k)W_{j4}(k)+W_{i4}(k)W_{j5}(k),\nonumber\\
&&W_{i2}(k)W_{j4}(k)+W_{i4}(k)W_{j2}(k)\Bigr ],
\end{eqnarray}
for the entries $\mu =6,...,15$.

In order to obtain the two-states energies, we need to diagonalize
simultaneously the Coulomb Hamiltonian and the crystal-field one.
We have already written the Coulomb Hamiltonian matrix in terms of
the triplets and the singlets. The next step is to express the
crystal-field Hamiltonian in terms of those. Omitting for brevity
the site index $k$, the crystal-field Hamiltonian matrix in the
triplet sector is
\end{multicols}
\begin{eqnarray}
\hspace*{-2em}
V_T^{\nil}=\left[\begin{array}{cccccccccc}
V_{22}\!+\!V_{55}&V_{23}&0&-V_{35}&V_{24}&0&-V_{45}&-V_{15}&0&V_{12}\\
V_{23}&V_{33}\!+\!V_{55}&0&V_{25}&V_{34}&V_{15}&0&0&-V_{45}&V_{13}\\
0&0&V_{11}\!+\!V_{44}&0&-V_{15}&-V_{34}&-V_{12}&V_{24}&-V_{13}&V_{45}\\
-V_{35}&V_{25}&0&V_{22}\!+\!V_{33}&0&V_{12}&V_{34}&V_{13}&-V_{24}&0\\
V_{24}&V_{34}&-V_{15}&0&V_{44}\!+\!V_{55}&0&V_{25}&0&V_{35}&V_{14}\\
0&V_{15}&-V_{34}&V_{12}&0&V_{11}\!+\!V_{33}&0&-V_{23}&-V_{14}&-V_{35}\\
-V_{45}&0&-V_{12}&V_{34}&V_{25}&0&V_{22}\!+\!V_{44}&V_{14}&V_{23}&0\\
-V_{15}&0&V_{24}&V_{13}&0&-V_{23}&V_{14}&V_{11}\!+\!V_{22}&0&V_{25}\\
0&-V_{45}&-V_{13}&-V_{24}&V_{35}&-V_{14}&V_{23}&0&V_{33}\!+\!V_{44}&0\\
V_{12}&V_{13}&V_{45}&0&V_{14}&-V_{35}&0&V_{25}&0&V_{11}\!+\!V_{55}
\end{array}\right]=V_T^t. \hspace*{-2cm}\nonumber\\
\end{eqnarray}
\begin{multicols}{2}
We are now in position to find the resolvent operator in  the
triplet sector. Denoting by $B$ the (unitary and real) matrix that
diagonalizes the triplet part of $H^{\text{cf}}+H^{\text{c}}$, and
by $E_{T}^{\mu}$ the corresponding eigenenergies,  we have
\begin{eqnarray}
\frac{1}{\Delta{\cal E}}\Psi_{T}^{\mu}(k;\sigma \sigma')=-\sum_{\mu '
}X_{T}^{k}(\mu ,\mu ')
\Psi_{T}^{\mu '}(k;\sigma \sigma'),\label{resolventT}
\end{eqnarray}
where $1/\Delta{\cal E}$ is the resolvent operator \cite{tak} and
\begin{eqnarray}
&&X_{T}^{k}(\mu,\mu ') = -\sum_{\mu_{1}
=1}^{10}\frac{B_{k}^{\nil}(\mu,\mu_{1} )B_{k}^{t}(\mu_{1} ,\mu
')}{E^{\mu_{1}}_{T}}.\label{xt}
\end{eqnarray}
One notes that since $B_k^{\nil}(\mu,\mu ')=B_k^t(\mu ',\mu)$, the matrices
$X_{T}^k$ satisfy
\begin{eqnarray}
X^{k}_{T}(\mu ,\mu ')=X_{T}^{k}(\mu ',\mu ).
\end{eqnarray}

Turning now to the singlet sector, we first find the crystal-field
Hamiltonian matrix of the singlets,
\begin{eqnarray}
V_S^{\nil}=\left[\begin{array}{ccc}V_{S1}^{\nil}&V_{S2}^{\nil}&V_{S3}^{\nil}\\
V_{S2}^t&V_{S4}^{\nil}&V_{S5}^{\nil}\\
V_{S3}^t&V_{S5}^t&V_{S6}^{\nil}\end{array}\right]=V_S^t,
\end{eqnarray}
where
\end{multicols}
\begin{eqnarray}
V_{S1}=\left[\begin{array}{cccccc}2V_{55}&0&0&0&0&\sqrt{2}V_{25}\\
0&2V_{22}&0&0&0&\sqrt{2}V_{25}\\
0&0&2V_{33}&0&0&0\\
0&0&0&2V_{44}&0&0\\
0&0&0&0&2V_{11}&0\\
\sqrt{2}V_{25}&\sqrt{2}V_{25}&0&0&0&V_{22}+V_{55}
\end{array}\right], \quad
V_{S2}=\left[\begin{array}{ccc} \sqrt{2}V_{15}&0&0\\
0&0&\sqrt{2}V_{12}\\
0&\sqrt{2}V_{34}&0\\
0&\sqrt{2}V_{34}&0\\
\sqrt{2}V_{15}&0&\sqrt{2}V_{12}\\
V_{12}&0&V_{15}
\end{array}\right], \nonumber
\end{eqnarray}
\begin{eqnarray}
V_{S3}=\left[\begin{array}{cccccc}0&\sqrt{2}V_{35}&0&0&\sqrt{2}V_{45}&0\\
0&0&\sqrt{2}V_{23}&0&0&\sqrt{2}V_{24}\\
0&\sqrt{2}V_{35}&\sqrt{2}V_{23}&\sqrt{2}V_{13}&0&0\\
\sqrt{2}V_{15}&0&0&0&\sqrt{2}V_{45}&\sqrt{2}V_{24}\\
\sqrt{2}V_{14}&0&0&\sqrt{2}V_{13}&0&0\\
0&V_{23}&V_{35}&0&V_{24}&V_{45}
\end{array}\right],\nonumber
\end{eqnarray}
\begin{eqnarray}
V_{S4}=\left[\begin{array}{ccc}V_{11}+V_{55}&0&V_{25}\\
0&V_{33}+V_{44}&0\\
V_{25}&0&V_{11}+V_{22}
\end{array}\right],\quad V_{S5}=\left[\begin{array}
{cccccc}V_{45}&V_{13}&0&V_{35}&V_{14}&0\\
V_{13}&V_{45}&V_{24}&V_{14}&V_{35}&V_{23}\\
V_{24}&0&V_{13}&V_{23}&0&V_{14}
\end{array}\right],\nonumber
\end{eqnarray}
\begin{eqnarray}
V_{S6}=\left[\begin{array}{cccccc}V_{11}+V_{44}&0&0&V_{34}&V_{15}&V_{12}\\
0&V_{33}+V_{55}&V_{25}&V_{15}&V_{34}&0\\
0&V_{25}&V_{22}+V_{33}&V_{12}&0&V_{34}\\
V_{34}&V_{15}&V_{12}&V_{11}+V_{33}&0&0\\
V_{15}&V_{34}&0&0&V_{44}+V_{55}&V_{25}\\
V_{12}&0&V_{34}&0&V_{25}&V_{22}+V_{44}
\end{array}\right].
\end{eqnarray}
\begin{multicols}{2}
\noindent Then we introduce  the ($15\times 15$) matrix $C$ that
diagonalizes $H^{\text{cf}}+H^{\text{c}}$ of the singlets and the
corresponding eigenenergies $E_{S}^{\mu}$. Analogously to Eq.
(\ref{xt}), it is convenient here to define as well
\begin{eqnarray}
&&X_{S}^{k}(\mu,\mu ') = -\sum_{\mu_{1}
=1}^{15}\frac{C_{k}^{\nil}(\mu,\mu _{1})C_{k}^{t}(\mu_{1} ,\mu '
)}{E^{\mu_{1}}_{S}},\label{xs}
\end{eqnarray}
which satisfies
\begin{eqnarray}
X^{k}_{S}(\mu ,\mu ')=X^{k}_{S}(\mu ',\mu ).
\end{eqnarray}
Analogously to Eq. (\ref{resolventT}) we have
\begin{eqnarray}
\frac{1}{\Delta{\cal E}}\Psi_{S}^{\mu}(k;\sigma \sigma')=-\sum_{\mu '
}X_{S}^{k}(\mu ,\mu ')
\Psi_{S}^{\mu '}(k;\sigma \sigma').
\end{eqnarray}

Collecting the results above, the intermediate two-particle states
of the perturbation expansion are now given in the form
\begin{eqnarray}
&&\frac{1}{\Delta{\cal
E}}d^{\dagger}_{ki\sigma}d^{\dagger}_{kj\sigma '}\nonumber\\
&=&\sum_{\mu\mu '=1}^{10}w_{T}^{\mu}(k;ij)X^{k}_{T}(\mu ,\mu
')\Psi_{T}^{\mu '}(k;\sigma\sigma ')\nonumber\\
&&+\sum_{\mu\mu '=1}^{15}w_{S}^{\mu}(k;ij)X^{k}_{S}(\mu ,\mu
')\Psi_{S}^{\mu '}(k;\sigma\sigma ').\label{cc1}
\end{eqnarray}
The final step involves transforming back $\Psi_{T}$ and
$\Psi_{S}$ into the $d$ operators. Consider, for example,
$\Psi_{T}^{\mu '}$  with $\mu '=\gamma_{1}\gamma_{2}$. Using Eqs.
(\ref{CFDIAG}) and (\ref{trisin}), we find (omitting the summation
notations for brevity)
\begin{eqnarray}
&&\Psi_{T}^{\gamma_{1}\gamma_{2}}(k;\sigma\sigma ')\nonumber\\
&=&\sqrt{\frac{1}{2}}\,W_{i_{1}\gamma_{1}}(k)W_{i_{2}\gamma_{2}}(k)\nonumber\\
&&\times \Bigl
(d^{\dagger}_{ki_{1}\sigma}d^{\dagger}_{ki_{2}\sigma '}
+d^{\dagger}_{ki_{1}\sigma '}d^{\dagger}_{ki_{2}\sigma }\Bigr
)\nonumber\\
&=&d^{\dagger}_{ki_{1}\sigma}d^{\dagger}_{ki_{2}\sigma
'}\nonumber\\
&&\times\sqrt{\frac{1}{2}}\Bigl
[W_{i_{1}\gamma_{1}}(k)W_{i_{2}\gamma_{2}}(k)-
W_{i_{2}\gamma_{1}}(k)W_{i_{1}\gamma_{2}}(k)\Bigr
]\nonumber\\
&=&d^{\dagger}_{ki_{1}\sigma}d^{\dagger}_{ki_{2}\sigma '}w^{\mu
'}_{T}(k;i_{1}i_{2}).
\end{eqnarray}
A similar calculation holds for the singlets. We therefore may
write
\begin{eqnarray}
\frac{1}{\Delta{\cal
E}}d^{\dagger}_{ki\sigma}d^{\dagger}_{ki'\sigma
'}&=&
Z_{k}^{\nil}(ii';i_{1}i_{2})d^{\dagger}_{ki_{1}\sigma}d^{\dagger}_{ki_{2}\sigma
'},
\end{eqnarray}
with
\begin{eqnarray}
Z_{k}(ii';i_{1}i_{2})&=& \sum_{\mu\mu
'=1}^{10}w^{\mu}_{T}(k;ii')X^{k}_{T}(\mu ,\mu ')w_{T}^{\mu
'}(k;i_{1}i_{2})\nonumber\\
&&+\sum_{\mu\mu '=1}^{15}w^{\mu}_{S}(k;ii')X^{k}_{S}(\mu ,\mu
')w_{S}^{\mu '}(k;i_{1}i_{2}).\nonumber\\
\end{eqnarray}
We note that since the $X$'s are symmetric, it follows that
\begin{eqnarray}
Z_{k}(ii';i_{1}i_{2})=Z_{k}(i_{1}i_{2};ii').\label{symz1}
\end{eqnarray}
Also, since $w_{S}^{\mu }(k;ii')=w_{S}^{\mu}(k;i'i)$ and
$w_{T}^{\mu}(k;ii')=-w_{T}^{\mu}(k;i'i)$, one has
\begin{eqnarray}
Z_{k}(ii';i_{1}i_{2})=Z_{k}(i'i;i_{2}i_{1}).\label{symz2}
\end{eqnarray}

\subsection{Perturbation expansion}

Our formal expressions of the various terms in the perturbation
expansion, Eqs. (\ref{part1}), (\ref{part2}), and (\ref{part3})
above, involve the projection operators $P_{mn}^0$ and
$S_{mn}=(1-P_{mn}^0)/\Delta{\cal E}$. Here we give their explicit
expressions in terms of the quantities derived in the first part
of this Appendix.

The projector onto the unperturbed ground-state space is
\begin{eqnarray}
P_{mn}^0=\sum_{\sigma\sigma'}d_{m0\sigma}^{\dag}
d_{n0\sigma'}^{\dag}\big|0\big>\big<0\big|d_{n0\sigma'}^{\nil}
d_{m0\sigma}^{\nil},
\end{eqnarray}
where $\big|0\big>$ denotes the vacuum state. Similarly, the
projector onto the Ti$^{3+}$ sector of the excited states is
\begin{eqnarray}
P_{mn}^1=\hspace*{-0.8em}\sum_{\sigma\sigma'\atop i_1i_2(\neq 00)}
\hspace*{-0.8em}d_{mi_1\sigma}^{\dag}d_{ni_2\sigma'}^{\dag}
\big|0\big>\big<0\big|d_{ni_2\sigma'}^{\nil}d_{mi_1\sigma}^{\nil}.
\end{eqnarray}
It follows that the resolvent operator  applied to $P_{mn}^1$ is
given by
\begin{eqnarray}
\frac{1}{\Delta{\cal E}}P_{mn}^1=
-\hspace*{-0.8em}\sum_{\sigma\sigma'\atop i_1i_2(\neq 00)}
\hspace*{-0.8em}\frac{1}{E_{i_1}\!+\!E_{i_2}}
d_{mi_1\sigma}^{\dag}d_{ni_2\sigma'}^{\dag}\big|0\big>
\big<0\big|d_{ni_2\sigma'}^{\nil}d_{mi_1\sigma}^{\nil}. \hspace*{-1cm}\nonumber \\
\end{eqnarray}
In a similar way, the projector onto the Ti$^{2+}$ sector of the
excited states is
\begin{eqnarray}
P_{mn}^2=\frac{1}{2}\sum_{k\sigma\sigma'}\Big[&&\sum_{\mu=1}^{10}
\Psi_{T}^{\mu}(k;\sigma\sigma ')\big|0\big>\big<0\big|
\Psi_{T}^{\mu \dag}(k;\sigma\sigma ')\nonumber \\
&&+\sum_{\mu=1}^{15}\Psi_{S}^{\mu}(k;\sigma\sigma ')
\big|0\big>\big<0\big|\Psi_{S}^{\mu \dag}(k;\sigma\sigma ') \Big],
\hspace*{-1cm}\nonumber \\
\end{eqnarray}
which gives, upon applying the resolvent operator,
\begin{eqnarray}
\frac{1}{\Delta{\cal E}}P_{mn}^2=-\frac{1}{2}\sum_{k\sigma\sigma'}\Big[&&\sum_{\mu \mu'}
X^k_T(\mu,\mu')\Psi_{T}^{\mu'}(k;\sigma\sigma ')\big|0\big>\nonumber \\[-2ex]
&& \hspace*{2cm}\times \big<0\big|
\Psi_{T}^{\mu \dag}(k;\sigma\sigma ')\nonumber \\[1ex]
&&+\sum_{\mu \mu'}X^k_S(\mu,\mu')\Psi_{S}^{\mu'}(k;\sigma\sigma ')
\big|0\big>\nonumber \\[-2ex]
&&\hspace*{2cm}\times \big<0\big|\Psi_{S}^{\mu \dag}(k;\sigma\sigma ') \Big].
\hspace*{-1cm}\nonumber \\
\end{eqnarray}
Hence, the combined resolvent and projection operator onto the
excited states is
\begin{eqnarray}
S_{mn}^{\nil}=\frac{1}{\Delta{\cal E}}(P_{mn}^1+P_{mn}^2).
\end{eqnarray}

Collecting these results, and expressing the products of $d$ in
terms of the ground-state spin operators [see Eqs.
(\ref{dStrafo})], one obtains the magnetic exchange couplings.
These are listed below.
\end{multicols}
\noindent {\it a. The Heisenberg couplings.} These are given by
\begin{eqnarray}
J_{mn}^{\nil}=-\sum_{i_{1}i_{2}}\big[t_{mn}^{i_{1}0}
Z_{m}^{\nil}(i_{1}0;0i_{2})t_{nm}^{0i_{2}}+
t_{nm}^{i_{1}0}Z_{n}^{\nil}(i_{1}0;0i_{2})t_{mn}^{0i_{2}}\big].
\end{eqnarray}
{\it b. The Moriya vectors.} These are given to first order in the
spin-orbit coupling,
\begin{eqnarray}
D_{mn}^{\alpha}=2i\lambda\sum_{i(\neq 0) \atop
i_{1}i_{2}}\frac{1}{E_i} \big\{&&L_{i0}^{\alpha}(m)
\big[t_{mn}^{i_{1}0}Z_{m}^{\nil}(i_{1}i;0i_{2})t_{nm}^{0i_{2}}+
t_{nm}^{i_{1}i}Z_{n}^{\nil}(i_{1}0;0i_{2})t_{mn}^{0i_{2}}\big] \nonumber\\[-2ex]
&&-L_{i0}^{\alpha}(n)
\big[t_{mn}^{i_{1}0}Z_{m}^{\nil}(i_{1}0;0i_{2})t_{nm}^{ii_{2}}+
t_{nm}^{i_{1}0}Z_{n}^{\nil}(i_{1}0;ii_{2})t_{mn}^{0i_{2}}\big]
\big\}.
\end{eqnarray}
{\it c. The symmetric anisotropies, and the $\lambda^2$ correction
of the Moriya vectors.} These terms are of second order in
$\lambda$, and have a more complicated structure. In order to
present them in a concised fashion, we write the left-hand-side of
Eq. (\ref{part3}) in the form
\begin{eqnarray}
{\mathbf{S}}_m^{\hspace*{0em}} A_{mn}^{\text{s}}
{\mathbf{S}}_n^{\hspace*{0em}}
\!+\!{\mathbf{D}}'_{mn}\big({\mathbf{S}}_m^{\hspace*{0em}}
\!\times\!{\mathbf{S}}_n^{\hspace*{0em}}\big)=
J'_{mn}{\mathbf{S}}_m^{\hspace*{0em}}{\mathbf{S}}_n^{\hspace*{0em}}
+\hspace*{-0.3em}\sum_{\alpha \beta l \atop {i(\neq 0)\atop
i'(\neq 0)}} \hspace*{-0.3em}\frac{1}{E_iE_{i'}}\,C_{mn}^{\alpha
\beta} (i,i',l) I_{mn}^{\alpha \beta} (l),\label{lhs}
\end{eqnarray}
where  $J'_{mn}$ and ${\mathbf{ D}}'_{mn}$ are the $\lambda^2$
corrections of the Heisenberg couplings and the Moriya vectors,
respectively. In Eq. (\ref{lhs}) $l$ enumerates the 4 spin
invariants, such that
\begin{eqnarray}
I_{mn}^{\alpha\beta}(1)&=&S_m^{\alpha} S_n^{\beta}, \quad
I_{mn}^{\alpha\beta}(2)=S_m^{\alpha} S_n^{\beta} +S_m^{\beta}
S_n^{\alpha}-\delta_{\alpha\beta}^{\nil}{\mathbf{S}}_m^{\hspace*{0em}}
{\mathbf{S}}_n^{\hspace*{0em}}, \nonumber \\
I_{mn}^{\alpha\beta}(3)&=&\delta_{\alpha\beta}^{\nil}{\mathbf{S}}_m^{\hspace*{0em}}
{\mathbf{S}}_n^{\hspace*{0em}} + \sum_{\gamma}
\epsilon_{\alpha\beta\gamma}^{\nil}
\big({\mathbf{S}}_m^{\hspace*{0em}}
\!\times\!{\mathbf{S}}_n^{\hspace*{0em}}\big)^{\gamma}, \quad
I_{mn}^{\alpha\beta}(4)=I_{mn}^{\beta \alpha}(3),
\end{eqnarray}
where $\epsilon_{\alpha\beta\gamma}$ is the totally antisymmetric
tensor. It remains to list the coefficients appearing in Eq.
(\ref{lhs}). These are given by
\begin{eqnarray}
J'_{mn}&=&-J_{mn}^{\nil} \frac{\lambda^2}{4} \sum_{ \alpha \atop
i(\neq 0)} \frac{1}{E_i^2}
 \big[ |L_{i0}^{\alpha}(m)|^2+|L_{i0}^{\alpha}(n)|^2 \big], \nonumber \\
C_{mn}^{\alpha \beta} (i,i',1)&=& 2 \lambda^2 \sum_{i_1i_2}
L_{i0}^{\alpha}(m)L_{0i'}^{\beta}(n)
\big[t_{mn}^{i_{1}0}Z_{m}^{\nil}(ii_{1};0i_{2})t_{nm}^{i'i_{2}}
+t_{nm}^{i_{1}0}Z_{n}^{\nil}(i'i_{1};0i_{2})t_{mn}^{ii_{2}}\nonumber \\[-1ex]
&&
\hspace*{3.5cm}-t_{mn}^{i_{1}0}Z_{m}^{\nil}(i_{1}0;i_{2}i)t_{nm}^{i'i_{2}}
-t_{nm}^{i_{1}0}Z_{n}^{\nil}(i_{1}0;i_{2}i')t_{mn}^{ii_{2}}
 \big],\nonumber \\
C_{mn}^{\alpha \beta} (i,i',2)&=&-\lambda^2
 \sum_{i_1 i_2}\Big\{ \frac{1}{2} L_{i0}^{\alpha}(m)
  L_{0i'}^{\beta}(m)\, \big[t^{i_10}_{mn}
  Z_m^{\nil}(i'i_1;i_2i)t^{0i_2}_{nm}+t^{i_1i'}_{nm}
  Z_n^{\nil}(i_10;0i_2)t^{ii_2}_{mn}\big]\nonumber \\[-1ex]
&& \hspace*{1.5cm} +\frac{1}{2} L_{i0}^{\alpha}(n)
L_{0i'}^{\beta}(n)
\big[t^{i_1i}_{mn}Z_m^{\nil}(i_10;0i_2)t^{i'i_2}_{nm}
+t^{i_10}_{nm}Z_n^{\nil}(ii_1;i_2i')t^{0i_2}_{mn}\big]\nonumber \\
&& \hspace*{1.5cm} - L_{i0}^{\alpha}(m) L_{0i'}^{\beta}(n)\,
 \big[t^{i_10}_{mn}Z_m^{\nil}(i_10;ii_2)t^{i'i_2}_{nm}+t^{i_10}_{nm}
 Z_n^{\nil}(i_10;i'i_2)t^{ii_2}_{mn}\big]\Big\},\nonumber \\
C_{mn}^{\alpha \beta} (i,i',3)&=&- \lambda^2 \sum_{i_1 i_2} \Big\{
\frac{1}{2}L_{i0}^{\alpha}(m)L_{0i'}^{\beta}(n)
\big[t_{mn}^{i_{1}i'}Z_{m}^{\nil}(i_{1}0;ii_{2})t_{nm}^{0i_{2}}
+t_{nm}^{i_{1}0}Z_{n}^{\nil}(i'i_{1};i_{2}0)
t_{mn}^{ii_{2}}\big]\nonumber \\[-1ex]
&& \hspace*{1.5cm}+ L_{i0}^{\alpha}(m)
 L_{i'i}^{\beta}(m) \, \big[t^{i_10}_{mn}
 Z_m^{\nil}(i_10;i'i_2)t^{0i_2}_{nm}+t^{i_10}_{nm}
 Z_n^{\nil}(i_10;0i_2)t^{i'i_2}_{mn}\big]\Big\}, \nonumber \\
C_{mn}^{\alpha \beta} (i,i',4)&=&- \lambda^2 \sum_{i_1 i_2}\Big\{
\frac{1}{2}L_{i0}^{\beta}(m)L_{0i'}^{\alpha}(n)
\big[t_{mn}^{i_{1}i'}Z_{m}^{\nil}(i_{1}0;ii_{2}) t_{nm}^{0i_{2}}
+t_{nm}^{i_{1}0}Z_{n}^{\nil}(i'i_{1};i_{2}0)
t_{mn}^{ii_{2}}\big]\nonumber \\[-1ex]
&& \hspace*{1.5cm}+ L_{i0}^{\alpha}(n) L_{i'i}^{\beta}(n) \,
\big[t^{i_10}_{mn}Z_m^{\nil}(i_10;0i_2)t^{i'i_2}_{nm}+t^{i_10}_{nm}
Z_n^{\nil}(i_10;i'i_2)t^{0i_2}_{mn}\big]\Big\}.
\end{eqnarray}

\begin{multicols}{2}

\setlength{\tabcolsep}{0.1cm}
\begin{table}
\caption{The parametrization of the unit cell (space group
$Pbnm$), modulo the lattice constants $a,b,c$.} \label{par}
\begin{tabular}{ll|l}
 La & & $(x_{\text{RE}},y_{\text{RE}},1/4),(1/2-x_{\text{RE}},1/2+y_{\text{RE}},1/4),$ \\
 & & $(-x_{\text{RE}},-y_{\text{RE}},3/4),(1/2+x_{\text{RE}},1/2-y_{\text{RE}},3/4)$ \\
 \hline
Ti & & $(0,1/2,0),(1/2,0,0),(0,1/2,1/2),(1/2,0,1/2)$ \\ \hline
O1 & & $(x_{\text{O1}},y_{\text{O1}},1/4),(1/2-x_{\text{O1}},1/2+y_{\text{O1}},1/4),$ \\
 & & $(-x_{\text{O1}},-y_{\text{O1}},3/4),(1/2+x_{\text{O1}},1/2-
 y_{\text{O1}},3/4)$ \\ \hline
O2 & & $(x_{\text{O2}},y_{\text{O2}},z_{\text{O2}}),(x_{\text{O2}},
y_{\text{O2}},1/2-z_{\text{O2}}),$ \\
 & & $(-x_{\text{O2}},-y_{\text{O2}},-z_{\text{O2}}),(-x_{\text{O2}},
 -y_{\text{O2}},1/2+z_{\text{O2}}),$ \\
 & & $(1/2-x_{\text{O2}},1/2+y_{\text{O2}},z_{\text{O2}}),$ \\
 & & $(1/2-x_{\text{O2}},1/2+y_{\text{O2}},1/2-z_{\text{O2}}),$ \\
 & & $(1/2+x_{\text{O2}},1/2-y_{\text{O2}},-z_{\text{O2}}),$ \\
 & & $(1/2+x_{\text{O2}},1/2-y_{\text{O2}},1/2+z_{\text{O2}})$
\end{tabular}
\end{table}
\renewcommand{\arraystretch}{1}
\setlength{\tabcolsep}{0cm}

\setlength{\tabcolsep}{0.1cm}
\begin{table}
\caption{The symmetries of the space group $Pbnm$.}
\label{sym}
\begin{tabular}{ll|l}
 Inversion centers & & Ti sites and centers of $ab$-planar Ti\\
& & plaquettes, i.\,e. $(0,0,0),(1/2,1/2,0),$ etc.  \\ \hline
 Mirror planes & & $z=\pm 1/4$\\ \hline
 Glide planes & & $x=\pm 1/4$, translation by $(0,1/2,0)$\\ \hline
 Screw axes & & Through the inversion centers, along the\\
& & $z$ axis, rotation by 180$^\circ$
\end{tabular}
\end{table}
\renewcommand{\arraystretch}{1}
\setlength{\tabcolsep}{0cm}

\renewcommand{\arraystretch}{1.3}
\setlength{\tabcolsep}{0.05cm}
\begin{table}
\caption{The static crystal field for Ti$^{3+}$ (site 1): Spectrum
and eigenstates in the orthorhombic basis for the $d$ orbitals,
see Eq. (\ref{dbasis}) and the following comment there. [The
eigenenergies $E_{i}$ and the matrix $W(1)$ used in conjunction
with Eq. (\ref{cftrafo}) are defined by the spectrum and the
coordinates of the eigenstates, respectively, as given in this
Table, where the first row of $W(1)$ is the coordinate vector of
the ground state, etc.]} \label{cf}
\begin{tabular}{rr|rrrrr}
--0.468 eV & & (--0.035,&0.016,& 0.770,& --0.035, & 0.636)\\ \hline
--0.259 eV & & (--0.052,&--0.397,& 0.088,&  0.911, &--0.049) \\ \hline
--0.239 eV & & (--0.407,&0.035,& --0.587,&  0.086, & 0.693) \\ \hline
 0.452 eV & & (\hspace*{0.5em}0.853,&0.315,& --0.197,&  0.221, & 0.290) \\ \hline
 0.515 eV & & (--0.319,&0.861,& 0.123,&  0.336, &--0.169) \\ \hline
Basis $\quad\;$& & $\big|xy\big>$,&$\big|2z^2\big>$,&$\big|yz\big>$,&$\big|xz\big>$,
&$\hspace*{-0.5em}\big|x^2\!-\!y^2\big>$
\end{tabular}
\end{table}
\setlength{\tabcolsep}{0cm}
\renewcommand{\arraystretch}{1}

\setlength{\tabcolsep}{0.05cm}
\renewcommand{\arraystretch}{1.3}
\begin{table}
\caption{The effective Ti--Ti hopping matrices for the $d$
eigen-orbitals of the crystal field from Table \ref{cf}; values
are given in eV. The rows and the columns are ordered beginning
with the ground state of the crystal field (index 0), continuing
with the first excited state (index 1), etc. $t_{13}$ is symmetric
due to the mirror plane at $z=1/4$.   As the matrices are given in
terms of the crystal-field eigen-basis, the dependence of the
hopping matrices on the bonds is particularly simple. [In
contrast, had we used the orthorhombic basis (\ref{dbasis}),
additional minus signs would have appeared in several entries.] }
\begin{tabular}{c}
Planar \\ \hline \\[-2.8ex]
$t_{12}^{\nil}=t_{16}^{t}=t_{25}^{\nil}=t_{65}^{t}=t_{34}^{\nil}
=t_{38}^{t}=t_{47}^{\nil}=t_{87}^{t}$\hspace*{1.9cm}\\[1ex]
$=\left[
\begin{array}{rrrrr}
-0.198& -0.155& -0.052& -0.022& 0.016\\
0.109& 0.133& 0.022& -0.089& 0.135\\
-0.114& 0.167& -0.188& -0.098& 0.193\\
-0.021& 0.088& -0.235& 0.579& -0.710\\
0.010& -0.019& -0.003& 0.089& -0.121
\end{array}
\right]_{\vspace*{1mm}}$ \\[-2.8ex] \\ \hline
Inter-planar \\ \hline \\[-2.8ex]
$t_{13}^{\nil}=t_{24}^{\nil}=t_{39}^{\nil}=t_{410}^{\nil}$\hspace*{5cm}\\[1ex]
$=\left[
\begin{array}{rrrrr}
0.178& \;\;\,0.047& -0.143& 0.010& 0.020\\
0.047& 0.244& 0.072& 0.135& 0.224\\
-0.143& 0.072& 0.146& -0.008& -0.057\\
0.010&  0.135& -0.008& -0.112& -0.312\\
0.020& 0.224& -0.057& -0.312& -0.812
\end{array}
\right]_{\vspace*{1mm}}$
\end{tabular}
\label{effhop}
\end{table}
\renewcommand{\arraystretch}{1}
\setlength{\tabcolsep}{0cm}

\setlength{\tabcolsep}{0.05cm}
\renewcommand{\arraystretch}{1.7}
\begin{table}
\caption{Model parameters of the calculation.}
\begin{tabular}{c}
Momenta of the effective ionic radius for Ti$^{3+}$ \\ \hline
$\big<r^2\big>=0.530\,\stackrel{\mbox{\tiny$\circ$}}
{\mbox{\small{A}}}\!\!\nil^2,\;\big<r^4\big>=0.554\,
\stackrel{\mbox{\tiny$\circ$}}{\mbox{\small{A}}}\!\!\nil^4$ \\ \hline
Slater integrals for Ti$^{2+}$\\ \hline
$F_2=8F_4/5=8.3\,\text{eV}$ \\ \hline
Spin-orbit parameter \\ \hline
$\lambda=18$\,meV\\ \hline
Slater-Koster parameters \\ \hline
$V_{pd\sigma}=-2.4$\,eV, $V_{pd\pi}=1.3$\,eV\\ \hline
Effective charge-transfer energies (Ti--Ti, Ti--O)\\ \hline
$U_{\text{eff}}=3.5$\,eV, $\Delta_{\text{eff}}=5.5$\,eV
\end{tabular}
\label{parameters}
\end{table}
\renewcommand{\arraystretch}{1}
\setlength{\tabcolsep}{0cm}

\renewcommand{\arraystretch}{1.3}
\setlength{\tabcolsep}{0.05cm}
\begin{table}
\caption{ The combined static and covalent crystal fields for
Ti$^{3+}$ (site 1): Spectrum and eigenstates in the orthorhombic
basis for the $d$ orbitals. The covalent contribution is
calculated for a TiO$_6$ cluster. The full eigenstates are linear
combinations of Ti$^{3+}$ states and Ti$^{2+}$ states (accompanied
by a $p$ hole on one of the oxygen sites). Here, only the
Ti$^{3+}$ parts of the five lowest eigenstates are shown,
corresponding to the states $\big|d^1\big>$ of Eq.
(\ref{clusterlc}).} \label{covcf}
\begin{tabular}{rr|rrrrr}
--0.663 eV & & (--0.029,&0.020,&0.778,&--0.023,&0.627)\\ \hline
--0.441 eV & & (\hspace*{0.5em}0.084,&--0.383,&0.200,&0.875,&--0.200) \\ \hline
--0.430 eV & & (--0.393,&--0.089,&--0.550,&0.279,&0.676) \\ \hline
 0.737 eV & & (\hspace*{0.5em}0.856,& 0.322,& --0.211,& 0.173,& 0.297) \\ \hline
 0.797 eV & & (--0.323,& 0.861,& 0.092,& 0.354,& --0.144) \\ \hline
Basis $\quad\;$& & $\big|xy\big>$,
&$\big|2z^2\big>$,&$\big|yz\big>$,&$\big|xz\big>$,&$\hspace*{-0.5em}\big|x^2\!-\!y^2\big>$
\end{tabular}
\end{table}
\setlength{\tabcolsep}{0cm}
\renewcommand{\arraystretch}{1}

\setlength{\tabcolsep}{0.05cm}
\renewcommand{\arraystretch}{1.3}
\begin{table}
\caption{Symmetries of the effective spin Hamiltonian due to the
space group. The relations among the anisotropic couplings
are abbreviated as follows. $(+,+,+)_{12}=(-,+,+)_{16}$ means
${\mathbf{D}}_{12}^{\hspace*{0em}}=(-D_{16}^x,D_{16}^y,D_{16}^z)$,
etc. Due to the mirror plane $z=1/4$, the inter-plane Moriya
vectors have vanishing $z$ components and the inter-plane symmetric
anisotropies have vanishing $yz$ and $xz$ entries. The transformation of
the symmetric anisotropies is characterized by the off-diagonal
coefficients $(A_{mn}^{yz},A_{mn}^{xz},A_{mn}^{xy})$ whereas the
diagonal coefficients are invariant in and between the planes,
respectively.}
\begin{tabular}{c}
Heisenberg couplings \\ \hline
$J_{12}=J_{16}=J_{25}=J_{65}=J_{34}=J_{38}=J_{47}=J_{87}$,\\
$J_{13}=J_{24}=J_{39}=J_{410}$ \\ \hline
Moriya vectors \\ \hline
$\;\,(+,+,+)_{12}=(-,+,+)_{16}=(+,-,-)_{25}=(-,-,-)_{65}$ \\
$\!=(-,-,+)_{34}=(+,-,+)_{38}=(-,+,-)_{47}=(+,+,-)_{87}$, \\
$\;(+,+,0)_{13}=(+,-,0)_{24}=(-,-,0)_{39}=(-,+,0)_{410}$ \\ \hline
Symmetric anisotropies \\ \hline
$\;\,(+,+,+)_{12}=(+,-,-)_{16}=(+,-,-)_{25}=(+,+,+)_{65}$ \\
$\!=(-,-,+)_{34}=(-,+,-)_{38}=(-,+,-)_{47}=(-,-,+)_{87}$, \\
$\;\,\;\;\,(0,0,+)_{13}=(0,0,-)_{24}=(0,0,+)_{39}=(0,0,-)_{410}$
\end{tabular}
\label{hamsym}
\end{table}
\renewcommand{\arraystretch}{1}
\setlength{\tabcolsep}{0cm}

\setlength{\tabcolsep}{0.05cm}
\renewcommand{\arraystretch}{1.3}
\begin{table}
\caption{The calculated single-bond spin couplings (in meV). The
Moriya vectors are given including the  corrections
$\mathbf{D}'_{mn}$, which are of order $\lambda^2$. The symmetric
anisotropies are given as
${\mathbf{A}}_{mn}^{\text{d}}=(A_{mn}^{xx},A_{mn}^{yy},A_{mn}^{zz})$
and
${\mathbf{A}}_{mn}^{\text{od}}=(A_{mn}^{yz},A_{mn}^{xz},A_{mn}^{xy})$
for the diagonal and off-diagonal entries, respectively.}
\begin{tabular}{c}
Heisenberg couplings \\ \hline
$J_{12}=24.616,\,J_{13}=19.416$ \\ \hline
Moriya vectors \\ \hline
${\mathbf{D}}_{12}=(3.254, -1.273, -1.286),
\,{\mathbf{D}}_{13}=(-2.886, 0.543,0)$\\ \hline
Symmetric anisotropies\\ \hline
${\mathbf{A}}_{12}^{\text{d}}=(0.188,0.066,0.037),
\,{\mathbf{A}}_{13}^{\text{d}}=(-0.039,-0.218,-0.190)$,\\
${\mathbf{A}}_{12}^{\text{od}}=(-0.035,-0.111,-0.088),
\,{\mathbf{A}}_{13}^{\text{od}}=(0,0,-0.074)$
\end{tabular}
\label{microscres}
\end{table}
\renewcommand{\arraystretch}{1}
\setlength{\tabcolsep}{0cm}

\setlength{\tabcolsep}{0.05cm}
\renewcommand{\arraystretch}{1.4}
\begin{table}
\caption{The macroscopic couplings of the sublattice
magnetizations in terms of the microscopic single-bond spin couplings.
For instance, we have $I_{12}=J_{12}$ but $I_{13}=J_{13}/2$.
This is because the coordination number of a Ti ion is 4 in the
planes and 2 between the planes.}
\begin{tabular}{c}
Isotropic couplings \\ \hline
$I_{12}=J_{12},\,I_{13}=\frac{1}{2}J_{13}$ \\ \hline
Dzyaloshinskii vectors \\ \hline
${\mathbf{D}}_{12}^{\text{D}}=(0,D_{12}^y,D_{12}^z),\,
{\mathbf{D}}_{13}^{\text{D}}=\frac{1}{2}{\mathbf{D}}_{13}^{\nil}$
\\ \hline Macroscopic symmetric anisotropies \\ \hline
$\Gamma_{12}^{\text{d}}={\mathbf{A}}_{12}^{\text{d}},\,
\Gamma_{12}^{\text{od}}=(A_{12}^{yz},0,0),\,\Gamma_{13}^{\nil}=\frac{1}{2}A_{13}^{\nil}$
\end{tabular}
\label{macrmicr}
\end{table}
\renewcommand{\arraystretch}{1}
\setlength{\tabcolsep}{0cm}

\setlength{\tabcolsep}{0.05cm}
\renewcommand{\arraystretch}{1.3}
\begin{table}
\caption{Symmetries of the magnetic  Hamiltonian due to the space
group. The relations for the anisotropic couplings are abbreviated
as in Table \ref{hamsym}. Due to the glide planes, the
Dzyaloshinskii vectors of the planar bonds have vanishing $x$ components,
and the respective symmetric anisotropies have vanishing $xz$ and $xy$
entries. Because of the mirror planes, the Dzyaloshinskii vectors
of the inter-planar bonds have vanishing $z$ components and the
respective symmetric anisotropies have vanishing $yz$ and $xz$ entries. }
\begin{tabular}{c}
Isotropic couplings \\ \hline $I_{12}=I_{34},\,I_{13}=I_{24}$ \\
\hline Dzyaloshinskii vectors \\ \hline
$(0,+,+)_{12}=(0,-,+)_{34}, \,(+,+,0)_{13}=(+,-,0)_{24}$  \\
\hline Macroscopic symmetric anisotropies \\ \hline
$(+,0,0)_{12}=(-,0,0)_{34}, \,(0,0,+)_{13}=(0,0,-)_{24}$
\end{tabular}
\label{macsym}
\end{table}
\renewcommand{\arraystretch}{1}
\setlength{\tabcolsep}{0cm}

\setlength{\tabcolsep}{0.15cm}
\renewcommand{\arraystretch}{1.3}
\begin{table}
\caption{All types of magnetic order which are allowed by  the
space group $Pbnm$. There are four possibilities, denoted by
$x^{\text{s}}$, $x^{\text{a}}$, $z^{\text{s}}$, and
$z^{\text{a}}$. They are allowed because the ordered state can be
symmetric or antisymmetric according to the glide plane $x=1/4$
and the mirror plane $z=1/4$, respectively.
The order in LaTiO$_3$ is of the first type. Here, $G_{x}$ denotes
G-type antiferromagnetic moment along $x$, $A_{y}$ denotes A-type
antiferromagnetic moment along $y$, and $F_{z}$ denotes
ferromagnetic moment along $z$.
The other possibilities involve also C-type ordering, e.\,g. $C_z$
for the $z$ components of the magnetic moments.
The magnetizations ${\mathbf{M}}_k$ of the sublattices are given
in terms of ${\mathbf{M}}_1$. $(+,\,+,\,+)_1=(-,\,+,\,+)_2$ means
${\mathbf{M}}_1^{\nil} =(-M_2^x,M_2^y,M_2^z)$, etc.}
\begin{tabular}{r|r|r|r}
  1. $x^{\text{s}},\,z^{\text{a}}\qquad$ & 2. $x^{\text{a}},\,z^{\text{a}}\qquad$
  & 3. $x^{\text{s}},\,z^{\text{s}}\qquad$ & 4. $x^{\text{a}},\,z^{\text{s}}\qquad$ \\
  \hline
$(+,\,+,\,+)_1$&$(+,\,+,\,+)_1$&$(+,\,+,\,+)_1$&$(+,\,+,\,+)_1$\\
=$(-,\,+,\,+)_2$&=$(+,\,-,\,-)_2$&=$(-,\,+,\,+)_2$&=$(+,\,-,\,-)_2$\\
=$(-,\,-,\,+)_3$&=$(-,\,-,\,+)_3$&=$(+,\,+,\,-)_3$&=$(+,\,+,\,-)_3$\\
=$(+,\,-,\,+)_4$&=$(-,\,+,\,-)_4$&=$(-,\,+,\,-)_4$&=$(+,\,-,\,+)_4$\\ \hline
 $G_x\, A_y \; F_z \,\;$ &  $A_x\,G_y\;C_z\,\;$&  $C_x\,F_y\;A_z\,\;$&  $F_x\,C_y\;G_z\,\;$
\end{tabular}
\label{alltypes}
\end{table}
\renewcommand{\arraystretch}{1}
\setlength{\tabcolsep}{0cm}

\setlength{\tabcolsep}{0.05cm}
\renewcommand{\arraystretch}{1.3}
\begin{table}
\caption{The structure of the magnetic order (the first
possibility of Table \ref{alltypes}), characterized by the
sublattice magnetizations ${\mathbf{M}}_k$ in the classical ground
state, in terms of the canting angles $\varphi$ and $\vartheta$.
Each of these angles is proportional to $\lambda$, the spin-orbit
parameter.}
\begin{tabular}{c}
$x$ components: G-type\\ \hline
$-M_1^x=M_2^x=M_3^x=-M_4^x=\cos \varphi \cos \vartheta $\\ \hline
$y$ components: A-type\\ \hline
$-M_1^y=-M_2^y=M_3^y=M_4^y=\sin \varphi \cos \vartheta $\\ \hline
$z$ components: ferromagnetic\\ \hline
$M_1^z=M_2^z=M_3^z=M_4^z=\sin \vartheta $
\end{tabular}
\label{cgs}
\end{table}
\renewcommand{\arraystretch}{1}
\setlength{\tabcolsep}{0cm}

\setlength{\tabcolsep}{0.05cm}
\renewcommand{\arraystretch}{1.3}
\begin{table}
\caption{The values of the macroscopic magnetic couplings
in meV, the resulting canting
angles of the magnetizations in the classical ground state, and
the resulting absolute values of the ordered moments (normalized
to 1\,$\mu_{\text{B}}$, i.\,e., without the reduction due to
quantum fluctuations and the spin-orbit coupling). The three
coefficients of the macroscopic symmetric anisotropies are taken
into account (see text).}
\begin{tabular}{c}
Isotropic couplings \\ \hline $I_{12}=24.616,\,I_{13}=9.708$ \\
\hline Dzyaloshinskii vectors \\ \hline
${\mathbf{D}}_{12}^{\text{D}}=(0, -1.273, -1.286),
\,{\mathbf{D}}_{13}^{\text{D}}=(-1.589, 0.271,0)$  \\ \hline
Macroscopic symmetric anisotropies \\ \hline
$\Gamma_{12}^{xx}=0.188,
\,\Gamma_{13}^{xx}=-0.020,\,\Gamma_{13}^{xy}=-0.037$ \\ \hline
Canting angles \\ \hline
 $\varphi=1.42^\circ,\,\vartheta=0.80^\circ$ \\ \hline
Ordered moments \\ \hline
${\mathbf{M}}=(0.9996,0.0248,0.0140)\,\mu_{\text{B}}$
\end{tabular}
\label{macroscres}
\end{table}
\renewcommand{\arraystretch}{1}
\setlength{\tabcolsep}{0cm}

\end{multicols}


\begin{thebibliography}{99}
\bibitem{norm} The $d$ orbitals given in Eq. (\ref{dbasis})
 are normalized when integrated over the unit sphere.
i.e.,  including the radial function $f(r)$, which obeys $\int
f(r)r^2dr=1$, we have the following representations in real space:
\\$\big<{\mathbf{r}} \big|xy\big>=f(r)\times \sqrt{15/4\pi}\,xy/r^2$,\\
$\big<{\mathbf{r}} \big|2z^2\big>=f(r)\times \sqrt{5/16\pi}\, (2z^2-x^2-y^2)/r^2$,\\
$\big<{\mathbf{r}} \big|yz\big>=f(r)\times \sqrt{15/4\pi}\,yz/r^2$,\\
$\big<{\mathbf{r}}\big|xz\big>=f(r)\times  \sqrt{15/4\pi}\,xz/r^2$,\\
$\big<{\mathbf{r}}\big|x^2-y^2\big>=f(r)\times  \sqrt{15/16\pi}\,(x^2-y^2)/r^2$.
\end{thebibliography}
\end{document}